\documentclass[reprint,amsmath,amssymb,aps,prb,
]{revtex4-1}

\usepackage{dcolumn}
\usepackage{bm}
\usepackage{xcolor}
\usepackage[normalem]{ulem}
\usepackage{epsf,latexsym,graphicx}
\usepackage{xr}
\usepackage{xcite}

\def \ba {\begin{eqnarray}}
\def \ea {\end{eqnarray}}
\def \ra {\rightarrow}

\begin{document}


\title{ Excitonic insulator emerging from semiconducting normal state in $1T$-TiSe$_{2}$ }

\author{Jin Mo Bok}
\altaffiliation[{\itshape Current address}:]
 { Department of Physics, Pohang University of Science and Technology (POSTECH), Pohang 37673, Korea.}
\affiliation{%
 Department of Physics, Sungkyunkwan University, Suwon 16419, Korea. }

\author{Jungseek Hwang}
\affiliation{%
 Department of Physics, Sungkyunkwan University, Suwon 16419, Korea. }

\author{Han-Yong Choi}%
\email{hychoi@skku.edu}
\affiliation{%
 Department of Physics, Sungkyunkwan University, Suwon 16419, Korea, \\
 Asia Pacific Center for Theoretical Physics, Pohang 37673, Korea.}




\date{\today}

\begin{abstract}

A new state of matter, an excitonic insulator (EI) state, was predicted to emerge from Bose-Einstein condensation of electron-hole pairs. Some candidate materials were suggested but it has been elusive to confirm its existence.
Recent works gave renewed support for the EI picture of the charge density wave (CDW) state below the critical temperature $T_c \approx 200 $ K of $1T$-TiSe$_{2}$. Yet, an important link to its establishment is to show that a majority fraction of the measured $T_c$ indeed follows from the Coulomb interaction alone, while a quantitative match of the $T_c$ may require assistance from the electron-lattice coupling. This will establish that the CDW is formed predominantly by the Coulomb interaction and help confirm the EI view for TiSe$_2$. Here, we provide such calculations by solving the exciton gap equation with material specific electronic structures. We obtain, with no fitting parameters, $T_c \approx 135 \pm 27$ K for the normal state gap of $E_g \approx 74 \pm 15$ meV. It seems that the calculated $T_c$ from Coulomb interaction gives a majority fraction of experimental $T_c$ for recently determined values of $E_g$. The measured doping dependence of $T_c$ was satisfactorily reproduced as well. Also in agreement with experiments are the same set of calculations of the photoemission spectroscopy and density of states. The semiconducting state above and EI below $T_c$ together should give a coherent picture of $1T$-TiSe$_2$.

\end{abstract}

\maketitle


\section{Introduction}

Coherent quantum states of excitons emerge out of a macroscopic number of electron-hole pairs bound by the Coulomb interaction as the band gap of semiconductors is reduced below the exciton binding energy.\cite{Mott1961philmag} They may Bose condense into a superfluid\cite{Snoke2002science} or an insulating electronic crystal.\cite{Jerome1967,Kohn1970rmp} The exciton superfluid or EI is a new state of matter which should have higher $T_c$ scale than the traditional Bardeen-Cooper-Schrieffer (BCS) or Bose-Einstein condensation (BEC) states due to their light mass and the strong binding energy.\cite{Halperin1968,Snoke2002science} These states may provide a new platform to investigate and utilize manifestations of macroscopic quantum phenomena.
Exciton condensation in electronic double layers, where indirect excitons form out of photo-generated electrons and holes residing in spatially-separated conducting layers, has been realized under strong magnetic field.\cite{Eisenstein2004nature}
More recently the exciton superfluidity transport was observed in quantized Hall regime in double bilayer graphenes at temperatures an order of magnitude
higher than previously observed in GaAs double layers.\cite{Liu2017naturephys,Li2017naturephys}
Despite these achievements there is a great need to identify materials in which an exciton condensate forms in equilibrium state without the applied field. TmSe$_{0.45}$Te$_{0.55}$, Ta$_2$NiSe$_5$, and $1T$-TiSe$_2$ have been investigated most intensely among candidate materials primarily because they have semiconducting/semimetallic electronic configuration with a small gap/overlap.

\begin{figure}
 \includegraphics[width=\linewidth]{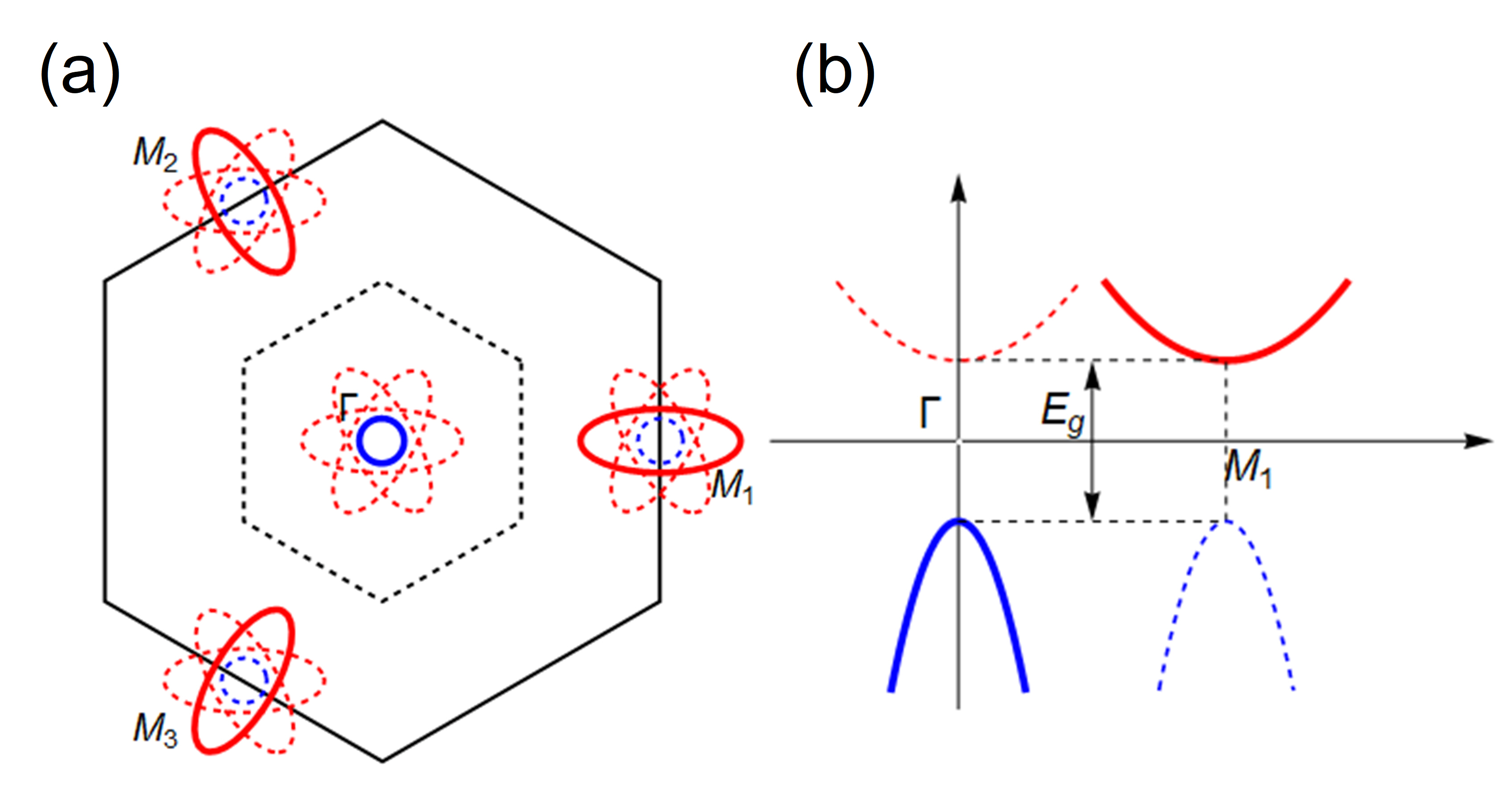}
\caption{(a) The 1st BZ of $1T$-TiSe$_2$ exhibiting the hole band (blue) around the $\Gamma$ and three electron bands (red) around $M_i$ points. The BZ is reduced in the CDW phase (black dashed line) and appear the backfolded dispersions (red and blue dashed lines). (b) The energy dispersion of the hole and electron bands along the $\Gamma-M_1$ direction. $E_g$ is the indirect bare gap between the electron and hole bands. The dashed lines are backfolded bands.}
\label{fermiband}
\end{figure}

The $1T$-TiSe$_2$ is a transition metal dichalcogenide semiconductor/semimetal of a layered structure with an indirect gap/overlap between the Se $4p$ hole band centered at the $\Gamma$ point and the Ti $3d$ electron bands around the $L$ ($M$ for single layer samples) points in the Brillouine zone (BZ) as shown in Fig.\ \ref{fermiband}.
An exciton, a bound state of an electron from the $M_i$ band and a hole from $\Gamma$ band in Fig.\ \ref{fermiband}(a), then has a nonzero net momentum and the inverse of the momentum sets a new length scale. Consequently, the exciton condensation is accompanied by a structural instability at the inverse momentum and makes a phase transition to CDW of $2a\times2a\times2c$ structure (where $a$ and $c$ are the lattice constants) below the critical temperature $T_c \approx 200$ K.
Early transport and angle-resolved photoemission spectroscopy (ARPES) measurements found that TiSe$_2$ has a small band gap/overlap and low numbers of carriers which led to speculations of the CDW as an EI.\cite{DiSalvo1976,Traum1978}
Also, the observation by high resolution ARPES of a very flat valence band dispersion near $\Gamma$ point and a large spectral weight transfer to backfolded bands in the CDW state seemed to be consistent with theoretical calculations on EI in TiSe$_2$.\cite{Monney2009,Cercellier2007}

Nature of the resulting CDW, however, is subtle to distinguish from other mechanisms involving lattice degrees of freedom alone like the Peierls instability\cite{george2000density} or Jahn-Teller distortion\cite{Rossnagel2002}.
An idea to differentiate them is to recall that the phonon mode corresponding to the CDW structure becomes soft at $T_c$ for electron-phonon coupled systems. Likewise, for a condensation of excitons of the momentum ${\bf Q}$ there should appear a softening in the Coulomb interaction, that is, plasma frequency softening at the ${\bf Q}$ as $T$ approaches $T_c$. The recent momentum resolved electron energy loss spectroscopy (EELS) experiments by Kogar $et~al.$ exactly observed this and gave a renewed support for the EI picture for $1T$-TiSe$_2$.\cite{Kogar2017}

On the other hand, it should be noted that the excitonic and lattice instabilities should appear simultaneously as reported by Kogar $et~al.$ because they have the same spatial symmetry. It means that the origin of the CDW should be rephrased as a quantitative question as to which between the instabilities contributes more to open the CDW gap and to what extent.
Then, a remaining theoretical step to a firm establishment of the EI in TiSe$_2$ is to check if a majority of the measured $T_c $ indeed follows from the Coulomb interaction alone in material specific calculations, while a quantitative match of the experimental $T_c$ may need assistance from the electron-lattice coupling.\cite{Porer2014} This will help establish that the CDW is formed predominantly by the Coulomb interaction, and confirm the EI view for TiSe$_2$.
Here, we provide such calculations by solving a BCS-like exciton gap equation \cite{Jerome1967,Monney2009} with experimentally determined electronic structures.

The employed exciton gap equation as given by Eqs.\ (\ref{eq:Coulomb}) and (\ref{eq:linear}) below is a mean-field approach which is reliable in the weak coupling BCS regime.
It also gives reliable $T_c$ in the strong coupling BEC regime
as was discussed by Bronold and Fehske.\cite{Bronold2006} From the perspective of BCS-BEC crossover theory this may seem surprising because the mean-field theory does not account for preformed excitons above $T_c$ on the semiconducting regime. However, the charge neutrality constraint in the present problem as given in (\ref{eq:S_neutral}) leads to a cancellation of the leading order corrections to the chemical potentials and forces $T_c$ on the semiconducting side to coincide with the BEC transition
temperatures for a noninteracting boson gas of excitons.\cite{Bronold2006}
This justifies the employment of the gap equation for $T_c$ determination in the semiconducting regime as well as the semimetallic one.

\begin{figure*}
\includegraphics[width=0.9\linewidth]{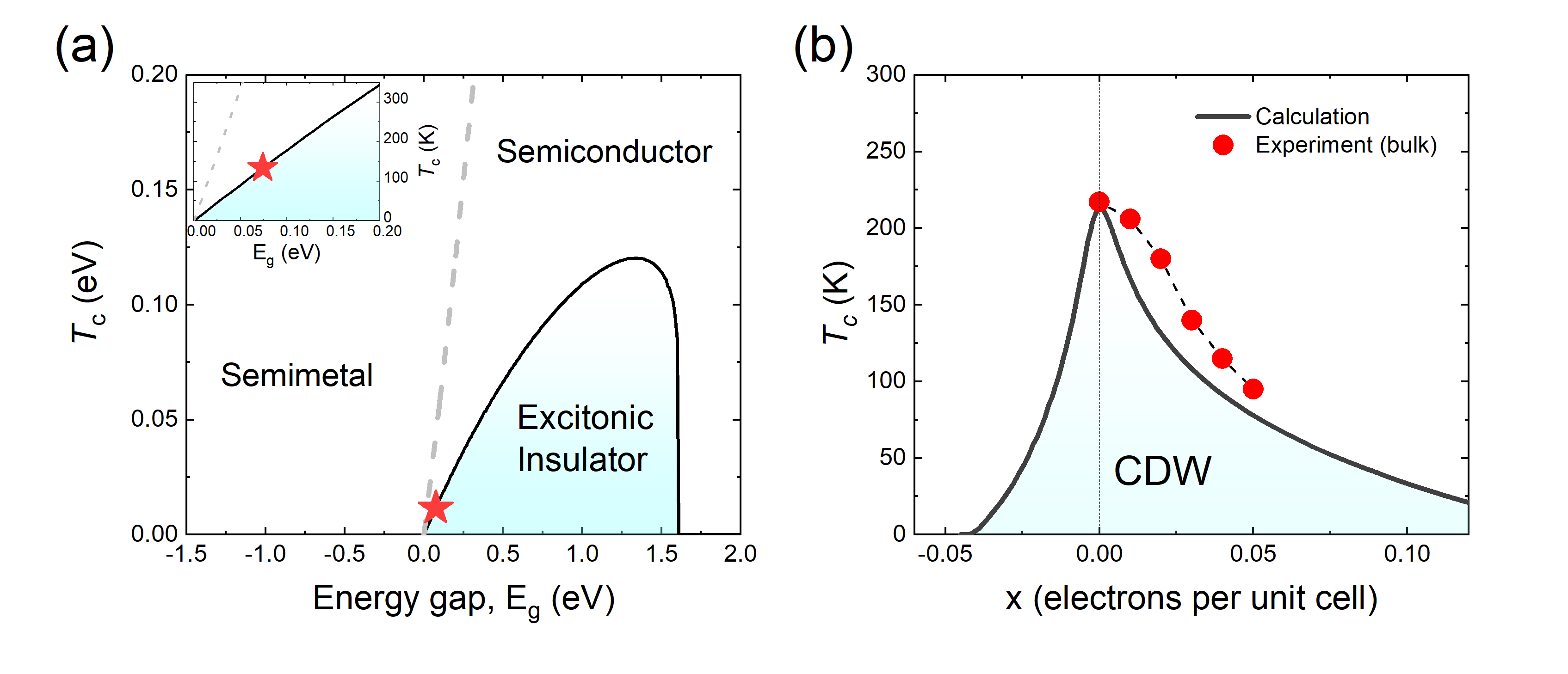}
\caption{
(a) The EI phase diagram of $1T$-TiSe$_{2}$ exhibiting $T_c$ as a function of $E_g$ for $\epsilon_\infty =3.0$. The grey dashed line indicates the border between semimetal and semiconductor phases. The red star on the $T_c$ curve shows the calculated $T_c$=135 K corresponding to the recent measurement of $E_g = 74$ meV.\cite{Watson2019PRL} The inset shows an enlargement of the region around the red star.
(b) The calculated doping dependence of $\mathit{T}_{c}$ for $E_g =0.12$ eV in comparison with experiments. The red circles represent measurements on pristine and Cu intercalated polycrystals.\cite{Morosan2006} }
\label{phasediag}
 \end{figure*}

\section{Formulation}

The calculations of the exciton order parameter $\Delta$ and the critical temperature $T_c$ presented here were performed by solving the exciton gap equation self-consistently by numerical iterations and finding the largest eigenvalue of the linearized gap equation. The gap equation is given by
 \begin{equation}
 \label{eq:gap}
\Delta_{i}(\mathbf{k})=-\sum_{\mathbf{k}'}\int_{-\infty}^{\infty}d\omega A_{\Delta_{i}}(\mathbf{k}',\omega)f(\omega)V_{s}(\mathbf{k}-\mathbf{k'}),
 \end{equation}
where
 \ba
\Delta_i (\mathbf{k}) = \langle a^\dagger (\mathbf{k}) b_i (\mathbf{k}) \rangle
 \ea
$(i=1,2,3)$ is the anomalous self-energy or the exciton order parameter formed between an electron from the $M_i$ band and a hole from around $\Gamma$, and $f$ is the Fermi-Dirac distribution function.
$A_\Delta$ is the anomalous spectral function given by
 \begin{eqnarray}\label{eq:ADelta}
&&A_{\Delta_{i}}(\mathbf{k}',\omega)=-\frac{1}{\pi}ImG_{\Delta_{i}}(\mathbf{k}',\omega) \\ \nonumber
&=&-\frac{1}{\pi}Im\frac{\Delta_{i}(\mathbf{k'})} {(\omega-\xi_{e,i})(\omega-\xi_{h})-\Delta_{i}^{2}(\mathbf{k'})-B_{i}},
 \end{eqnarray}
where $\omega$ should be understood as $\omega+i\delta$, with $\delta$ an infinitesimal.
The $\xi_{e,i}$ and $\xi_{h}$ are the dispersions for the $i$-th electron band and hole band given by
 \ba \label{eq:band}
\xi_{e,i}(\mathbf{k}) &=& \varepsilon^0_{e,i}(\mathbf{k})-\mu+E_{g}+\Sigma_{e,i},
 \nonumber \\
\xi_{h}(\mathbf{k}) &=& \varepsilon^0_{h}(\mathbf{k})-\mu+\Sigma_{h},
 \ea
where $E_g$ is the bare energy gap between the electron and hole bands and $\mu$ is the chemical potential. $\varepsilon^0_{e,i}$ and $\varepsilon^0_h$ are the bare dispersions of electron and hole bands, respectively, given by Eq.\ (\ref{eq:baredispersion}), and $\Sigma_{e,i}$ and $\Sigma_h$ are the corresponding self-energy given by Eq.\ (\ref{eq:S_Sigma}) in the Appendix \ref{sec:formulation}.

$B_i$ is the term arising from the coupling between the three electron bands given by
 \begin{equation}\label{eq:B}
B_{i}(\omega,\mathbf{k'})=\sum_{j\ne i}\Delta_{j}^{2}(\mathbf{k'})\frac{\omega-\xi_{e,i}(\mathbf{k'})} {\omega-\xi_{e,j}(\mathbf{k'})},
 \end{equation}
with $\mathit{i}$ and $\mathit{j}$ the electron band indices.
A derivation of the exciton gap equation with the explicit electronic structure of TiSe$_2$ is outlined in the Appendix \ref{sec:formulation}. A similar formulation was presented by Monney $et~al$ before.\cite{Monney2009}

For the kernel of the gap equation (\ref{eq:gap}), we employed the two dimensional (2D) Thomas-Fermi screened Coulomb interaction as the effective interaction between charge carriers in TiSe$_2$.
\ba \label{eq:Coulomb}
V_{s}(\mathbf{q})=\frac{2\pi e^{2}}{\epsilon_{\infty}(q+q_{s})} .
 \ea
$\epsilon_\infty$ is the inplane component of the background dielectric tensor of the 3D TiSe$_2$ of layered structure and $q_s$ is the screening wavenumber defined in Eq.\ (\ref{eq:S_qs}).
Deduction of this form for TiSe$_2$ from the screened Coulomb interaction of layered structures, $V_{s}^{3D} (\mathbf{q},q_z)$ of Eq.\ (\ref{eq:Coulomb3D}), is discussed in Appendix \ref{sec:screened}.

In the limit of $T\ra T_c$, $\Delta_i$, and $B_i$ of Eq.\ (\ref{eq:B}) $\ra 0$, and the exciton gap equation of (\ref{eq:gap}) is reduced to the linearized gap equation \cite{Bronold2006}
\begin{equation} \label{eq:linear}
\Delta(\mathbf{k})=\int_{BZ}\frac{d\mathbf{k'}}{(2\pi)^{2}}V_{s}(\mathbf{k}-\mathbf{k'}) \frac{f(\xi_{h}(\mathbf{k'}))-f(\xi_{e}(\mathbf{k'}))} {\xi_{e}(\mathbf{k'})-\xi_{h}(\mathbf{k'})} \Delta(\mathbf{k'}).
\end{equation}
The order parameters $\Delta_i $ are related by a rotational symmetry and satisfy the same form of the gap equation as can be seen from Eq.\ (\ref{eq:gap}). The solution of $T_c$ is independent of the band index $i$ and it was dropped out.

Eq.\ (\ref{eq:linear}) formally looks identical to the BCS gap equation. This similarity, however, holds only if the electron and hole have the equal masses, their dispersions are direct (negative) gap semimetals and satisfy $\xi_e ({\bf{k}}) = - \xi_h (-\bf{k})$ for one electron and one hole bands. In real materials these stringent requirements are necessarily not met and the celebrated Cooper logarithmic divergence is washed away. Consequently, the exciton instability is no longer a universal phenomenon in the BCS sense but becomes number comparison problems of the exciton binding energy being larger than the band gap. Material specific calculations are called for as discussed below.

\section{Exciton insulator critical temperature}

\begin{figure*}
\includegraphics[width=0.9\linewidth]{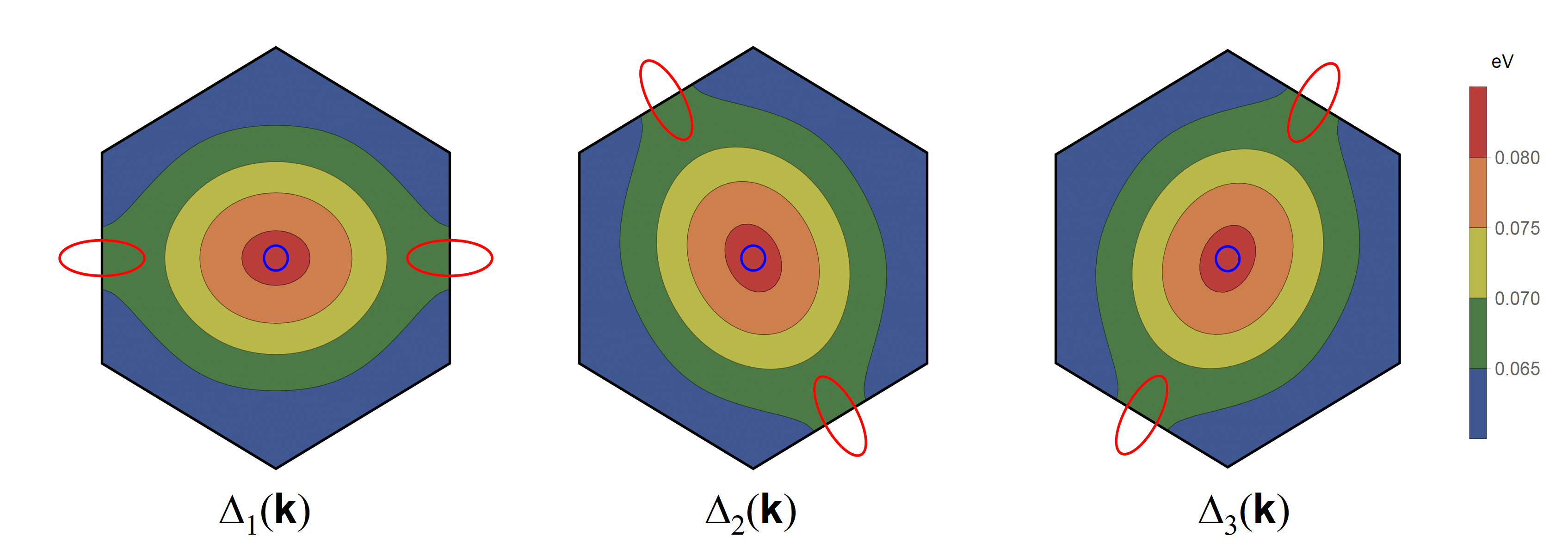}
\caption{The momentum dependence of the computed order parameter $\Delta_i(\mathbf{k})$ within the 1st BZ. The blue and red lines are the hole and electron bands, respectively. $\Delta_i(\mathbf{k})$ takes the largest value at the BZ center and is elongated along $\Gamma-M_i$ direction. }
\label{Delta}
\end{figure*}

It is crucial to determine the quantities $q_{s}$, $\Sigma_{i}$ and $\mu$ self-consistently as we did here by solving Eqs.\ (\ref{eq:S_qs}), (\ref{eq:S_neutral}), and (\ref{eq:S_Sigma}) via numerical iterations, for given temperature $T$, the background dielectric constant $\epsilon_\infty$, and normal gap $E_g$. The self-consistency ensures the $T_c$ from the exciton gap equation is reliable in the BEC as well as BCS regime as mentioned in the last paragraph in the introduction section.
With thus determined quantities, we write Eq.\ (\ref{eq:linear}) in a matrix form by discretizing the wavevector ${\bf k}$ within an irreducible BZ as
 \ba \label{matrix}
\Delta({\bf k}_i) =\sum_j M(i,j) \Delta({\bf k}_j).
 \ea
We took the wavevector ${\bf k}$ as an array of $48 \times 48$ $k$-points within the hexagonal 1st BZ as explained in Appendix \ref{sec:Tc}. Eq.\ (\ref{matrix}) is just an eigenvalue problem with the eigenvalue set to 1.
Above $T_c$ all the eigenvalues are smaller than 1. Then, $T_{c}$ is given by the temperature where the largest eigenvalue of the matrix $M$ of Eq.\ (\ref{matrix}) becomes 1.
This sets up a route to determine the EI critical temperature $T_c$ microscopically with no fitting parameters.

$\mathit{T}_{c}$ of $1T$-TiSe$_2$ was calculated as the normal state gap $E_g$ is varied taking the background dielectric constant $\epsilon_{\infty} = 3.0$. This gives the calculated $T_c$ once the normal gap $E_g$ is experimentally established. The determination of $\epsilon_{\infty}$ from the published experiments data and {\it ab initio} calcualtions is explained in Appendix \ref{sec:epsilon}.
The resulting phase diagram in the $E_g-T$ plane is shown in Fig.\ \ref{phasediag}{(a)}. As can be seen from the figure,
$T_c \approx 135 \pm 27 $ K for the normal gap $E_g \approx 74 \pm 15 $ meV (marked by the red star) recently determined by utilizing the photon-energy dependent ARPES as we discussed about the normal state gap in the paragraph after next.\cite{Watson2019PRL}
The measured CDW critical temperature is $T_c \approx 202$ K from the neutron diffraction measurement on bulk crystals,\cite{DiSalvo1976} and on monolayers $\mathit{T}_{c}\approx 232$ K from ARPES \cite{PChen2015} and $ \approx 240$ K from Raman experiments.\cite{Goli2012} The calculated $T_c \approx 135 \pm 27 $ K is smaller than the measured $T_c \approx 200$ K from bulk samples but is larger than half of it.

We also checked robustness of the results by calculating $T_c$ from the 3D gap equation of (\ref{eq:linear3D}) and (\ref{eq:Coulomb3D}). The screened Coulomb interaction is weakened in 3D which decreases $T_c$. On the other hand, for a fixed $E_g$, the minimum gap between $\Gamma$ and $L$ points in 3D BZ, the $k_z$ dispersion of the bands widens the average energy gap between the electron and hole bands, which should increases $T_c$ around $E_g \approx 74$ meV as can be inferred from Fig.\ \ref{phasediag}{(a)}. This overcomes the decrease from the weakened Coulomb interaction and causes a net increase to $T_c \approx 150 $ K for $E_g \approx 74$
meV as shown in Fig.\ \ref{fig:Tc_3D} in Appendix \ref{sec:Tc}. Notwithstanding uncertainty in the normal state nature, it seems that the CDW formation in TiSe$_2$ is mainly by the Coulomb interaction. This should provide a much needed theoretical support for the EI picture of the CDW state of TiSe$_2$.

We wish to point out that the screened Coulomb interaction in monolayers of transition metal dichalcogenides is better described by the Rytova-Keldysh potential because of the dielectric contrast between a monolayer and surrounding substrates or vacuum.\cite{Chernikov2014} The parameters of the potential need to be determined before quantitative comparisons are made between experimental $T_c$ and calculations for monolayers.

The EI phase in the low temperature limit appears for $0<E_g < E_X $, where $E_X $ is the exciton binding energy. $E_X =1.6$ eV for $\epsilon_\infty =3.0 $. It is given by
 \ba \label{EX}
E_X \approx 2 m_r c^2 \left( \frac{\alpha}{\epsilon_\infty} \right)^2 -\frac{\hbar^2 \overline{\Gamma M}^2}{2 M},
 \ea
where $\alpha = e^2/\hbar c$, $\overline{\Gamma M}$ the distance between $\Gamma$ and $M$ in the BZ, and $m_r$ and $M$ are, respectively, the reduced mass and total mass of an electron and a hole of an exciton, as discussed in Appendix \ref{sec:Tc}.
The phase diagram for direct gap EI including its suppression in the semimetallic region was discussed by Bronold and Fehske.\cite{Bronold2006}
The EI phase may be suppressed in the semimetallic region by the indirectness of the gap, asymmetric mass ratio between electron and hole, or the number of conduction bands being larger than 1.
Eq.\ (\ref{eq:linear}), which determines the EI phase diagram, is different from the superconducting gap equation as discussed above and is dependent on material specific parameters including
$\epsilon_\infty$. Consequently, the EI phase diagrams presented here and in \cite{Bronold2006} look quite different from the presumed EI phase diagram \cite{Lu2017} or from the BCS-BEC crossover of superconductors.

The normal state must be semiconducting for the EI to be a viable picture for the CDW phase of TiSe$_2$ as can be manifestly seen in Fig.\ \ref{phasediag}{(a)}. Yet, nature of the normal state is still controversial. Some of previous works reported semimetallic behaviors.\cite{DiSalvo1976,Rossnagel2002,Kusmartseva2009,PChen2015,jaouen2019}
ARPES measurements showed that the electron band at the $L$ point touches or crosses the Fermi level in the normal state and moves down to a lower energy as $T$ is decreased.\cite{Rossnagel2002,PChen2015} It is consistent with metallic behaviour of the $T$ dependence of resistivity.\cite{DiSalvo1976}
On the other hand, many spectroscopic and transport experiments reported semiconducting normal state.\cite{Kidd2002,Kolekar2018,Watson2019PRL}
Watson $et~al.$, utilizing the photon energy-dependent ARPES, presented results supporting the semiconducting normal state, $E_g \approx 74 \pm 15$ meV.\cite{Watson2019PRL} They also showed that the ``passenger'' states from the unhybridized Ti $d$ conduction bands and the valence band around the $A$ point remain decoupled from the CDW instability and provide the metallic behavior by accommodating extra charges from various extrinsic factors like crystal imperfections, Se vacancies, excess Ti and residual iodine.
Novello $et~al$ also noticed that the semimetallic behavior in previous works actually came from crystal imperfections which can be controlled by new synthesis techniques, and that the imperfections are not related to the CDW phase formation.\cite{Novello2015} Measurements on samples from the new synthesis techniques exhibited the anticipated insulating low temperature behaviors.\cite{Campbell2019,Moya2019}

To strengthen the arguments for the EI picture for TiSe$_2$, we consider the doping dependence of $T_c$ by repeating the calculations as we vary the chemical potential.
The doping dependence, $\mathit{T}_{c} (x)$, can be obtained by calculating $T_c$ and the doping concentration $x$ as a function of the chemical potential $\mu$. Because the doping increases the screening wavenumber $q_s$, the Coulomb binding energy is weakened, and $\mathit{T}_{c} (x)$ should be a decreasing function of $x$.
Fig.\ \ref{phasediag}(b) shows that the calculated $\mathit{T}_{c} (x)$ (black solid line) is in good agreement with the electron doping experiments.
The $T_c$ suppression rate by hole doping is steeper than the electron doping as might be expected because the conduction band has higher density of states (DOS). The $T_c(x)$ result also suggests that pristine TiSe$_2$ to exhibit intrinsic behavior might be prepared by hole/electron doping such that the samples have a maximum $T_c$ with doping.

The pressure dependence of $T_c$ may be considered similarly.
Pressure induces more overlap between the electron and hole bands and a decrease of the gap. This leads to more screening and weakened interaction, which suppresses the $T_c$ within the EI picture. This behavior was indeed observed experimentally.\cite{Kusmartseva2009}

\begin{figure*}
\includegraphics[width=0.9\textwidth]{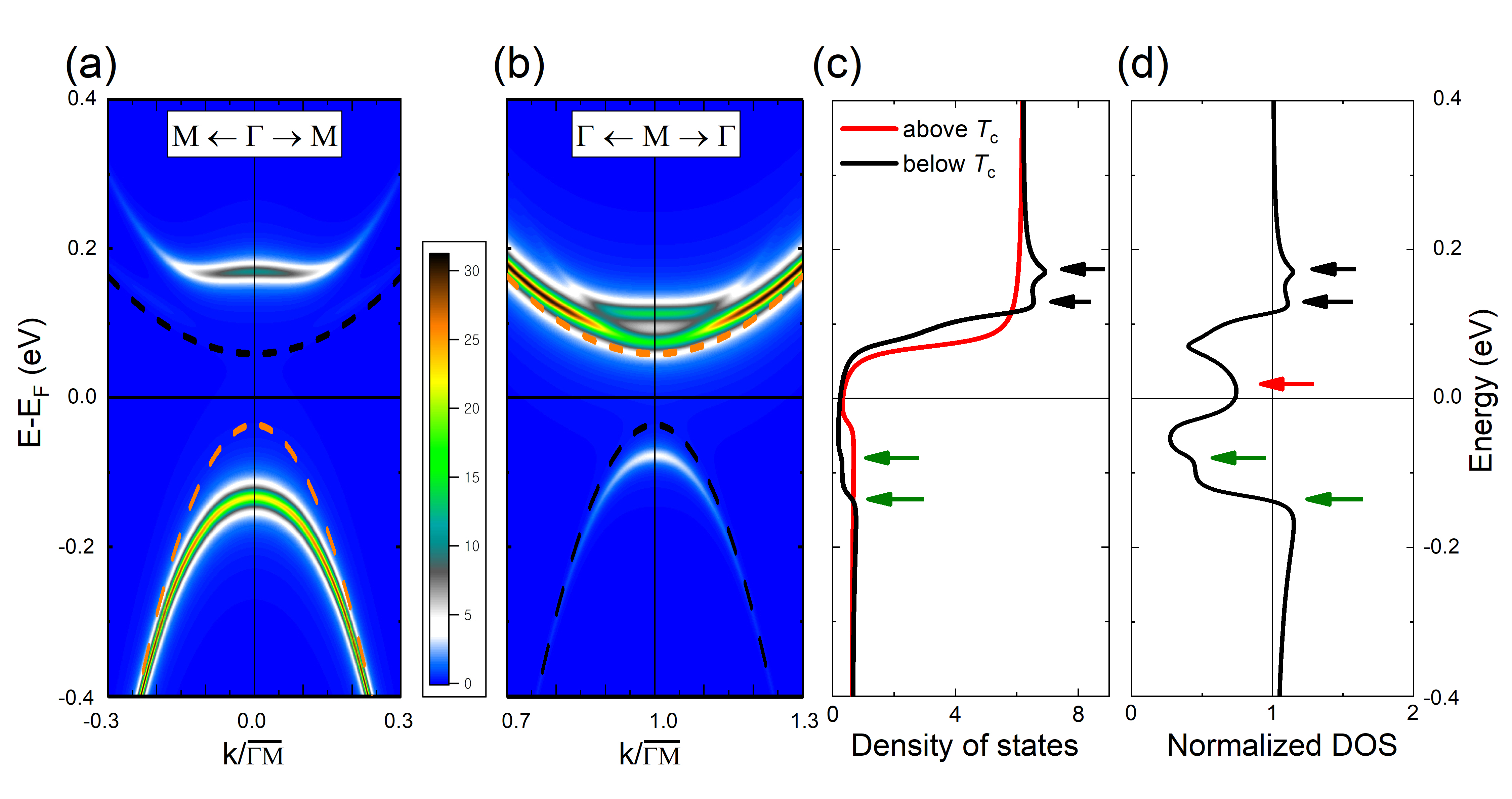}
\caption{(a),(b) The calculated spectral function around $\Gamma$ and $M$ points. The orange dashed lines represent the hole ((a)) and electron ((b)) dispersions in the normal phase and the black dashed lines show their backfolded ones. The high intensity curves near the orange lines are the main dispersions in the EI state.
(c) DOS in EI (black line) and normal (red) states. The black and green arrows indicate the coherence peaks due to flat dispersions of spectral function. These peaks are split because of the momentum dependence of $\Delta_i ({\bf k})$.
(d) The normalized DOS, where the zero bias peak (marked by the red arrow) shows up because of the normal state gap. }
 \label{arpes}
\end{figure*}

\section{Spectroscopic features}

We now turn to the spectroscopic features of the ARPES and DOS below $T_c$.
In the low temperature regime deep in the CDW state, the full gap equation of Eq.\ (\ref{eq:gap}) must be solved self-consistently to obtain $\Delta_i (\mathbf{k})$. All three gap order parameters $\Delta_i (\mathbf{k})$ are coupled in the full gap equation unlike the linearized one, and they are calculated simultaneously. We assumed they are real for simplicity with no consideration of the possible chirality of the order parameter phase.
The sum over ${\bf k'}$ in Eq.\ (\ref{eq:gap}) was performed using the fast Fourier transform with a $48 \times 48$ $k$-point mesh within the hexagonal 1st BZ as is explained in Appendix \ref{sec:dos}. The indirectness of the gap was properly taken into consideration in this process.
The computed $\Delta_i ({\bf k})$ is shown in the 1st BZ in Fig.\ \ref{Delta}.

With the determined $\Delta_i ({\bf k})$ as shown in Fig.\ \ref{Delta}, the spectral functions of the hole and electron bands which ARPES measure are then given as below.
 \ba\label{spectral}
A_{e,i}(\mathbf{k},\omega) &=& -\frac{1}{\pi}Im\frac{\omega-\xi'_{h}-B_{i}/(\omega-\xi_{e,i})} {(\omega-\xi_{e,i})(\omega-\xi'_{h})-\Delta_{i}^{2}(\mathbf{k})-B_{i}}, \nonumber \\
A_{h}(\mathbf{k},\omega) &=& - \frac{1}{\pi}Im\frac{\omega-\xi'_{e,3}} {(\omega-\xi'_{e,3})(\omega-\xi_{h})-\Delta_{3}^{2}(\mathbf{k})-B'_{3}},\nonumber \\
 \ea
where $\xi'$ are the backfolded dispersions of the corresponding electron and hole bands.
The DOS is obtained by summing the spectral function $A(\mathbf{k},\omega)$ over the wavevector {\bf k} by
 \begin{equation}
D(\omega)=\sum_{\mathbf{k}} \left[ \sum_{i=1,2,3} A_{e,i}(\mathbf{k},\omega)+ A_{h}(\mathbf{k},\omega) \right].
 \label{dos}
 \end{equation}

The renormalized dispersions in the EI states show up in high intensity as shown in Fig.\ \ref{arpes}(a) and (b).
The calculated spectra are in good agreement with Monney $\textit{et al}$.\cite{Monney2009} The most significant distinction is that the exciton order parameter with the full momentum dependence was calculated self-consistently from the screened Coulomb interaction in the present work.
The conduction and hole dispersions are hybridized in the EI state. Around, say, the $M_1$ point, the $\xi_{e,1}$ out of the triplet conduction bands and the hole dispersion $\xi_h$ are coupled and shifted up and down, respectively, and show up with high intensity as shown in (a) and (b). Their backfolded bands also clearly show up.
The renormalized conduction band is shift up from $\xi_{e,1}$ and shows a flat dispersion around the $M_1$ point.
The other two electron dispersions remain little affected. Consequently, they overlap with the bare dispersions (the orange dashed lines) and their backfoldings have a vanishing intensity.

Some of these features may be better studied by the the scanning tunneling spectroscopy/microscopy (STS/STM) experiments because ARPES can not probe the unoccupied states. The tunneling conductance $(dI/dV)$ is proportional to DOS given by Eq.\ (\ref{dos}).
The calculated DOS above and below $T_c$ are shown in Fig.\ \ref{arpes}(c). The coherence peaks become split due to the momentum dependence of the exciton order parameter; two step-like structure in the negative bias (green arrows) and two peak structure in the positive bias (black arrows) regime.
Interesting features may be revealed more clearly in the normalized conductance of $(dI/dV)_{CDW}/(dI/dV)_{normal}$ as shown in Fig.\ \ref{arpes}(d). Most important is the clear zero bias peak of the width of the normal gap as marked by the red arrow. This is because the normal state is semiconducting with the gap $E_g$ and the EI has a bigger gap of $\sqrt{E_g^2 + 3(2 \Delta_\Gamma)^2 }$ or $\sqrt{E_g^2 + 3(2\Delta_M)^2 }$.
Systematic analysis of DOS is not available in the literature. But, the recent STS experiments of Kolekar $et~ al$.\cite{Kolekar2018} (reproduced in Appendix \ref{sec:dos}) do exhibit the anticipated zero bias peak, which is consistent with the semiconducting normal state. High resolution STS data in a wide bias range will be informative.

\section{Summary and concluding remarks}

We presented in this paper a theoretical investigation of the view that the charge density wave below $T_c \approx 200$ K is an excitonic insulator phase in $1T$-TiSe$_2$. The idea was to check if the calculated $T_c$ from the Coulomb interaction alone accounts for a majority fraction of the experimental $T_c$. If so, it implies that the observed CDW is predominantly formed by the Coulomb interaction and that the CDW is an EI phase. Two crucial physical parameters for a quantitative determination of $T_c$ were the background dielectric constant $\epsilon_\infty$ and the normal state gap $E_g$. We determined $\epsilon_\infty \approx 3.0 $ by combining the published reflectance spectrum and EELS measurements up to 30 eV. If we take $E_g \approx 74 \pm 15 $ meV as a representative value, we obtained $T_c \approx 135 \pm 27 $ K. This accounts for majority fraction of the measured $T_c \approx 200 $ K from bulk crystals, while quantitative match of the measured $T_c$ may need assistance from other interactions like the electron-lattice coupling and the Jahn-Teller effect. It seems that the EI is perhaps a convincing view of the CDW state of $1T$-TiSe$_2$.

The EI view was furthermore reinforced by calculating the doping dependence of the critical temperature $T_c (x)$ and the spectroscopic features in the low temperature limit. These calculations were done with the same set of parameters as the $T_c$ calculations of undoped samples and were in agreement with experimental observations. We note again that the EI phase emerges only out of a semiconducting normal state (positive $E_g$). The semiconducting state above $T_c$ and EI below together should give a coherent picture of $1T$-TiSe$_2$.

The semimetallic or semiconducting normal state behavior turns out to be a crucial facet of underlying physics which may help uncover nature of CDWs of disparate mechanism.
The underlying physics of the Peierls instability or Jahn-Teller distortion is to lift degeneracy or to reduce DOS at the Fermi level by a CDW distortion to lower the free energy. This implies that a metallic DOS at the Fermi level is favored for the Peierls instability or Jahn-Teller distortion to be operative. On the other hand, the excitons may condense in the metallic or insulating states in, respectively, BCS or BEC regimes. Therefore, a CDW out of a semiconducting normal state is expected to be an excitonic nature. This physical anticipation is indeed borne out to be the case for TiSe$_2$ by detailed material specific calculations in this paper.
We anticipate our approach to be applied to wider class of problems and materials where the exciton condensation is yet to be explored.


\appendix

\section{ Formulation of the exciton gap equation for $1T$-TiSe$_{2}$ \label{sec:formulation} }

To describe the excitonic insulator state of $1T$-TiSe$_{2}$, we considered three Ti $3d$-derived electron bands around $M_i$ and a Se $4p$-derived hole band around $\Gamma$ in the BZ with statically screened Coulomb interaction. The Hamiltonian is written as
\begin{eqnarray}
H&=&H_{0}+W \\
H_{0}&=&\sum_{\mathbf{k}}\varepsilon_{h}(\mathbf{k})a^{\dagger}(\mathbf{k})a(\mathbf{k})
 + \sum_{\mathbf{k},i}\varepsilon_{e,i}(\mathbf{k})b_i^{\dagger}(\mathbf{k})b_i (\mathbf{k}) \\
W&=&\frac{1}{2}\sum_{\mathbf{q},i}\rho_{a}(\mathbf{q})V_{s}(\mathbf{q})\rho_{b,i}(\mathbf{-q}),
\end{eqnarray}
where the density operators are
\begin{eqnarray}
\rho_{a}(\mathbf{q})&=&\sum_{\mathbf{k}}a^{\dagger}(\mathbf{k}+\mathbf{q})a(\mathbf{k}), \\
\rho_{b,i}(\mathbf{q})&=&\sum_{\mathbf{k}}b_{i}^{\dagger}(\mathbf{k}+\mathbf{q})b_{i}(\mathbf{k}).
\end{eqnarray}
Here $\varepsilon_{h}$ and $\varepsilon_{e,i}$ are the hole and electron dispersions and $i$ is the electron band index. $V_s(\mathbf{q})$ is the Thomas-Fermi screened Coulomb interaction in 2D given in Eq.\ (\ref{eq:Coulomb}) in the main text.
$q_{s}$ is the screening wavenumber given by
 \begin{equation}\label{eq:S_qs}
q_{s}=\frac{2\pi e^{2}}{\epsilon_{\infty}} \frac{\partial n}{\partial\mu},
\end{equation}
where $\mu$ is the chemical potential and $n$ is the sum of the densities of three electron and a hole bands,
 \ba
 n(\mu,T)= \sum_{i=1,2,3} n_{e,i}(\mu,T) +n_{h}(\mu,T),
 \label{density}
 \ea
with
 \ba
n_{e,i}(\mu,T)=2\int_{BZ}\frac{d\mathbf{k}}{(2\pi)^{2}} f(\xi_{e,i}) , \nonumber \\
n_{h}(\mu,T)=2\int_{BZ}\frac{d\mathbf{k}}{(2\pi)^{2}} f(-\xi_{h}) ,
 \ea
where the factor of 2 is due to the spin degeneracy.

The charge neutrality condition is
 \ba \label{eq:S_neutral}
\sum_{i=1,2,3} n_{e,i}(\mu,T) = n_h (\mu,T).
 \ea
Deviations from the charge neutrality is represented by the doping concentration $x$ which can be calculated by varying the chemical potential $\mu$. $x$ is given by
 \ba
 x(\mu,T)= \sum_{i=1,2,3} n_{e,i}(\mu,T) -n_{h}(\mu,T),
 \label{doping}
 \ea
for given temperature $T$ and $\mu$.

We employ the four component operator $\Psi_{\mathbf{k}}$ in Nambu notation.
\begin{equation}
\Psi_{\mathbf{k}}^{\dagger}=\left( b_{1}^{\dagger}(\mathbf{k}),b_{2}^{\dagger}(\mathbf{k}), b_{3}^{\dagger}(\mathbf{k}), a^{\dagger}(\mathbf{k}) \right), ~~
\Psi_{\mathbf{k}}=\left(
\begin{array}{cccc}
b_{1}(\mathbf{k}) \\
b_{2}(\mathbf{k}) \\
b_{3}(\mathbf{k}) \\
a(\mathbf{k})
\end{array} \right).
\end{equation}
The $4\times4$ matrix renormalized Green's function $\hat{G}$ is defined as
\begin{eqnarray}
\hat{G}(\mathbf{k},\tau)=-\langle T_{\tau}\Psi_{\mathbf{k}}(\tau)\Psi_{\mathbf{k}}^{\dagger}(0)\rangle ,
\end{eqnarray}
with the bare Green's function $\hat{G_0}$ and the self-energy $\hat{\Sigma}$ given as
\begin{equation}
\hat{G}_{0}^{-1}(\mathbf{k},ip)=ip-
\left(\begin{array}{cccc}
\varepsilon_{e,1}(\mathbf{k}) & 0 & 0 & 0 \\
0 & \varepsilon_{e,2}(\mathbf{k}) & 0 & 0 \\
0 & 0 & \varepsilon_{e,3}(\mathbf{k}) & 0 \\
0 & 0 & 0 & \varepsilon_{h}(\mathbf{k})
\end{array}\right),
\end{equation}
\begin{equation}
\hat{\Sigma}(\mathbf{k},ip)=
\left(\begin{array}{cccc}
\Sigma_{e,1}(\mathbf{k},ip) & 0 & 0 & \Delta_{1}(\mathbf{k},ip) \\
0 & \Sigma_{e,2}(\mathbf{k},ip) & 0 & \Delta_{2}(\mathbf{k},ip) \\
0 & 0 & \Sigma_{e,3}(\mathbf{k},ip) & \Delta_{3}(\mathbf{k},ip) \\
\Delta_{1}(\mathbf{k},ip) & \Delta_{2}(\mathbf{k},ip) & \Delta_{3}(\mathbf{k},ip) & \Sigma_{h}(\mathbf{k},ip)
\end{array}\right).
\end{equation}
The renormalized Green's function is given by Dyson's equation as
\begin{eqnarray}
\hat{G}^{-1}(\mathbf{k},ip) = \hat{G}_{0}^{-1}(\mathbf{k},ip)-\hat{\Sigma}(\mathbf{k},ip) \nonumber \\ \nonumber\\
 = \left( \begin{array}{cccc}
ip - \xi_{e,1}(\mathbf{k}) & 0 & 0 & -\Delta_{1}(\mathbf{k},ip) \\
0 & ip - \xi_{e,2}(\mathbf{k}) & 0 & -\Delta_{2}(\mathbf{k},ip) \\
0 & 0 & ip - \xi_{e,3}(\mathbf{k}) & -\Delta_{3}(\mathbf{k},ip) \\
-\Delta_{1}(\mathbf{k},ip) & -\Delta_{2}(\mathbf{k},ip) & -\Delta_{3}(\mathbf{k},ip) & ip - \xi_{h}(\mathbf{k})
\end{array}\right) \nonumber \\
\end{eqnarray}
in the Matsubara frequency.

The normal state dispersions are
 \ba \label{eq:dispersion}
\xi_{e,i}(\mathbf{k}) &=& \varepsilon^0_{e,i}(\mathbf{k})-\mu+E_{g}+\Sigma_{e,i},
 \nonumber \\
\xi_{h}(\mathbf{k}) &=& \varepsilon^0_{h}(\mathbf{k})-\mu+\Sigma_{h},
 \ea
as given previously in the main text.
The diagonal self-energies $\Sigma_{e,i}$ and $\Sigma_h$ are given by
\ba\label{eq:S_Sigma}
\Sigma_{e,i}= -2\sum_{\mathbf{k}}V_{s}(\mathbf{k})f(\xi_{i}(\mathbf{k})), \nonumber \\
\Sigma_{h}= 2\sum_{\mathbf{k}}V_{s}(\mathbf{k})f(-\xi_{h}(\mathbf{k})).
\ea
$\Sigma$ and $q_s$ were also determined self-consistently by keeping the charge neutrality.\cite{Bronold2006}
The renormalized band gap is
 \ba
 \Bar{E_g} = E_g+\Sigma_e-\Sigma_h.
 \ea

The 2D bare dispersions within the hexagonal 1st BZ were taken as
 \ba \label{eq:baredispersion}
\varepsilon^0_{e,1}(k_{x},k_{y})&=& \frac{\hbar^{2}}{2m_{e,l}}\left(k_{x}-\overline{\Gamma M}\right)^{2}+\frac{\hbar^{2}}{2m_{e,s}}k_{y}^{2},
 \nonumber \\
\varepsilon^0_{e,2}(k_{x},k_{y})&=& \frac{\hbar^{2}}{2m_{e,l}}\left(\frac{1}{2}\left(k_{x}+\frac{\overline{\Gamma M}}{2}\right)- \frac{\sqrt{3}}{2}\left(k_{y}-\frac{\sqrt{3}}{2}\overline{\Gamma M}\right)\right)^{2} \nonumber \\
 &+& \frac{\hbar^{2}}{2m_{e,s}}\left(\frac{\sqrt{3}}{2}\left(k_{x}+\frac{\overline{\Gamma M}}{2}\right) + \frac{1}{2}\left(k_{y}-\frac{\sqrt{3}}{2}\overline{\Gamma M}\right)\right)^{2},
 \nonumber \\
\varepsilon^0_{e,3}(k_{x},k_{y})&=& \frac{\hbar^{2}}{2m_{e,l}}\left(\frac{1}{2}\left(k_{x}+\frac{\overline{\Gamma M}}{2}\right)+\frac{\sqrt{3}}{2}\left(k_{y}+\frac{\sqrt{3}}{2}\overline{\Gamma M}\right)\right)^{2} \nonumber \\
 &+& \frac{\hbar^{2}}{2m_{e,s}}\left(\frac{\sqrt{3}}{2}\left(k_{x}+\frac{\overline{\Gamma M}}{2}\right) -\frac{1}{2}\left(k_{y}+\frac{\sqrt{3}}{2}\overline{\Gamma M}\right)\right)^{2},
 \nonumber \\
\varepsilon^0_{h}(k_{x},k_{y})&=& -\frac{\hbar^{2}}{2m_{h}}\left(k_{x}^{2}+k_{y}^{2}\right),
 \ea
for the three elliptic electron bands at $M_i $ points and a hole band at $\Gamma$,
where $\overline{\Gamma M}$ is the distance between $\Gamma$ and $M$ points in the reciprocal space.
We took
\begin{eqnarray}
m_{e,l}=3.46\times\textrm{free electron mass}, \nonumber \\
m_{e,s}=1.38\times\textrm{free electron mass}, \nonumber \\
m_{h}=0.63\times\textrm{free electron mass},
\end{eqnarray}
for the effective masses of the conduction and hole bands.\cite{PChen2015,CChen2018}

\section{ The screened Coulomb interaction for TiSe$_2$ of layered structure \label{sec:screened} }

\begin{figure}
\includegraphics[width=0.5\textwidth]{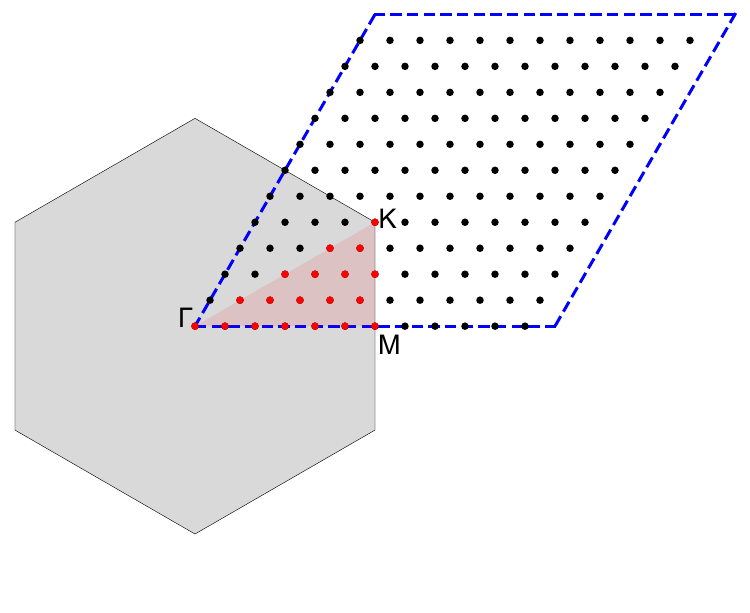}
\caption{ The gray hexagonal area represents 1st BZ of TiSe$_2$. It can be converted into the blue dash rhombus of the same area. The red and black dots in the blue dashed rhombus are the selected $k$-points for FFT. The red dots are the $k$-points in the irreducible BZ (pale red triangle) for $T_c$ calculations.}
 \label{k_points}
\end{figure}

We calculated the critical temperature $T_c$ of the exciton condensation by solving the linearized gap equation (\ref{eq:linear}). The inputs are the material specific electronic structure and the effective screened Coulomb interaction $V_s$. Let us first consider $V_s$.

Recall that the solution to the anisotropic Poisson's equation with the in-plane and and out-of-plane dielectric constants $\epsilon_\parallel $ and $\epsilon_\perp $ is given
 \ba \label{aniso_Coulomb}
V({\bf r},z) = \frac{e}{\sqrt{\epsilon_\parallel \epsilon_\perp }} \frac{1}{ \sqrt{ r^2 +\frac{\epsilon_\parallel}{ \epsilon_\perp } z^2} }
 \ea
as may be verified by a direct substitution to the Poisson's equation.
 \ba
\epsilon_\parallel \left( \frac{\partial^2}{\partial x^2} + \frac{\partial^2}{\partial y^2} \right)
 +\epsilon_\perp \frac{\partial^2}{\partial z^2} = - 4\pi e \ \delta({\bf r},z).
 \ea
${\bf r}$ is the 2D inplane vector. Take the distance between neighboring layers as $c$. A scaling to ${\tilde z} = \sqrt{\epsilon_\parallel / \epsilon_\perp } z $ in Eq.\ (\ref{aniso_Coulomb}) reveals that this is a Coulomb potential of a charge $e$ in an isotropic medium of a dielectric constant $\sqrt{\epsilon_\parallel \epsilon_\perp }$ in the ${\bf r},~ {\tilde z} $ space.

We then follow the well studied results of the dielectric function of layered structures in an isotropic medium.\cite{Visscher1971,Shung1986} It is given in the reciprocal space by
 \ba
\epsilon ({\bf q},{\tilde q}_z) = \sqrt{\epsilon_{\infty}^{\parallel} {\epsilon_{\infty}^{\perp}} } + v_q \chi \frac{\sinh(qd)}{\cosh(qd)- \cos( {\tilde q}_z d)},
 \ea
where $d= \sqrt{ \epsilon_{\infty}^{\parallel} / \epsilon_{\infty}^{\perp} } c$, 
${\tilde q}_z = \sqrt{ \epsilon_{\infty}^{\perp} / \epsilon_{\infty}^{\parallel} } q_z$,
$v_q = 2\pi e^2/q$, and $\chi$ is the response function.
The effective screened Coulomb interaction is
 \ba \label{eq:Coulomb3D}
V_s^{3D} & & ({\bf q},{\tilde q}_z) = \frac{2\pi e^2}{\sqrt{\epsilon_{\infty}^{\parallel} {\epsilon_{\infty}^{\perp}} } q}
\left[ \frac{ \cosh(qd)- \cos( {\tilde q}_z d) } {\sinh(qd)} + \frac{q_s}{q} \right]^{-1} , \nonumber \\
 \ea
where $q_s = 2\pi e^2 \chi/\sqrt{\epsilon_{\infty}^{\parallel} {\epsilon_{\infty}^{\perp}} }$.

The exciton gap equation in 2D of (\ref{eq:linear}) would be written in 3D as
\ba \label{eq:linear3D}
 \Delta(\mathbf{k},{\tilde k}_z) &=& \frac{1}{(2\pi)^2} \int_{BZ} d\mathbf{k'} \frac{c}{2\pi} \int d{\tilde k}'_z V_{s}^{3D}( {\bf k}-{\bf k}',{\tilde k}_z -{\tilde k}'_z) \nonumber \\
 &\times& \frac{f(\xi_{h}({\bf k}',{\tilde k}'_z))-f(\xi_{e}({\bf k}',{\tilde k}'_z))} {\xi_{e}({\bf k}',{\tilde k}'_z)-\xi_{h}({\bf k}',{\tilde k}'_z)} \Delta({\bf k}',{\tilde k}'_z).
\ea
For cases where the $z$-direction dispersion can be neglected in (\ref{eq:linear3D}) as in strongly anisotropic materials, we may ignore the ${\tilde k}_z$ dependence of $\Delta(\mathbf{k},{\tilde k}_z)$ and perform the ${\tilde k}'_z$ integral. Then, Eq.\ (\ref{eq:linear3D}) is reduced to Eq.\ (\ref{eq:linear}) where
 \ba
V_s({\bf q}) &=& \frac{c}{2\pi} \int_{-\pi/d}^{\pi/d} d{\tilde q}_z V_s^{3D} ({\bf q},{\tilde q}_z) \nonumber \\
 &=& \frac{2\pi e^2}{ \epsilon_{\infty}^{\parallel} q} \frac{\sinh(qd)}
 {\sqrt{ \left[\cosh(qd) + \sinh(qd) \frac{q_s}{q} \right]^2 -1} } ,
 \ea
using the relation of $\int d {q}_z V({ q}_z) = \int d {\tilde q}_z V({\tilde q}_z) $. Note that the effective background dielectric constant $ \sqrt{\epsilon_{\infty}^{\parallel} {\epsilon_{\infty}^{\perp} } } $ of 3D layered materials is changed to $ \epsilon_{\infty}^{\parallel}$ because of the $\sqrt{\epsilon_{\infty}^{\parallel} / \epsilon_{\infty}^{\perp} } $ factor in the integrand in the $q_z$ integral.
Make expansion on the small $q_s d$ and use $\tanh( q_{min} d)= 1.0$, where $q_{min} =\overline{\Gamma M}$ corresponds to the wavevector of the minimum gap between $\Gamma $ and $M$ points. The $q_{min} d \gg 1$ corresponds to a weak coupling limit which reduces the Coulomb interaction of a layered structure to that of decoupled single layers.
We obtain
 \ba
V_s({\bf q}) = \frac{2\pi e^2}{ \epsilon_{\infty}^{\parallel}} \frac{1}{q + q_s }.
 \ea
This shows that the effective background dielectric constant in the 2D modelling of the screened Coulomb interaction is given by the inplane component of the background dielectric tensor of 3D layered materials.

\section{ Calculation of $\mathit{T}_{c}$ \label{sec:Tc} }

There are a few characteristics of the electronic structure of candidate materials relevant for the exciton condensation: (a) direct or indirectness of band gap, (b) the normal gap size $E_g$, (c) the number of electron bands and hole bands, (d) mass asymmetricity, the ratio of hole to electron masses $\beta = {m_h}/{m_e}$, and (e) the background dielectric constant. All these material specific electronic structures were incorporated in the calculations of the critical temperature $T_c$ of the exciton condensation.

To calculate $T_c$, the linearized gap equation was written in a matrix form by discretizing the wavevector ${\bf k}$ within 1st BZ as given in Eq.\ (\ref{matrix}) in the main text. The discretization of the wavevector ${\bf k}$ was taken as the $48 \times 48$ points as shown in Fig.\ \ref{k_points}.
For calculation efficiency, ${\bf k}_i$ points were selected within the irreducible BZ (red dots in red area) with the weight of each point properly taken into account.
This produces $\sim 1/12$ times smaller matrix size to diagonalize.
Eq.\ (\ref{matrix}) is just an eigenvalue problem with the eigenvalue set to 1. For temperatures above the critical temperature $T_c$ all the eigenvalues are smaller than 1. $T_{c}$ is given by the temperature where the largest eigenvalue of the matrix $M$ of Eq.\ (\ref{matrix}) becomes 1.
This sets up a route to determine the EI critical temperature $T_c$ microscopically with the electronic structure properties like $\epsilon_\infty$ and $E_g$ and with no fitting parameters.

\begin{figure*}
\includegraphics[width=0.9\textwidth]{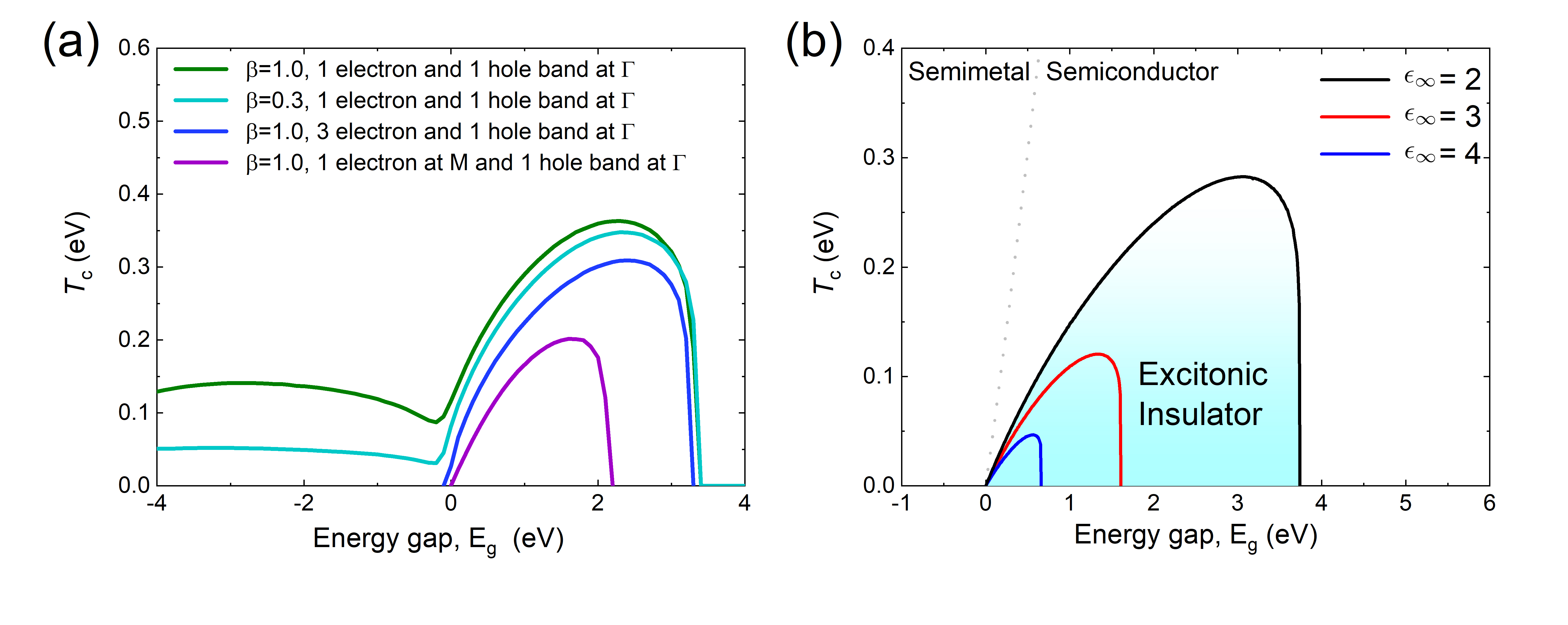}
\caption{ {(a)} The critical temperature $T_c$ of excitonic instability as a function of the energy gap for four different characteristic configurations of electronic structure. {(b)} shows $T_c$ vs.\ $E_g$ for TiSe$_2$ for different choices of the background dielectric constant $\epsilon_\infty$. }\label{S_Tc_model}
\end{figure*}

Fig.\ \ref{S_Tc_model}{(a)} shows the critical temperature $T_c$ of EI in 2D for four different cases of the characteristics.
The calculated EI phase on the negative $E_g$ region strongly depends on the material parameters, while it is more robust on the positive $E_g$ side. All of the characteristic configurations affect the suppression of EI phase on the negative $E_g$ region. In particular, EI phase on the negative $E_g$ region is completely suppressed for indirect gap configurations or the number of conduction bands being greater 1. The exciton condensation is not expected in the semimetallic cases for TiSe$_2$. Fig.\ \ref{S_Tc_model}{(b)} shows $T_c$ vs.\ $E_g$ for TiSe$_2$ for different choices of the background dielectric constant $\epsilon_\infty $ = 2, 3, and 4. $T_c$ is suppressed as $E_g$ is increased.

\begin{figure*}
\includegraphics[width=0.6\textwidth]{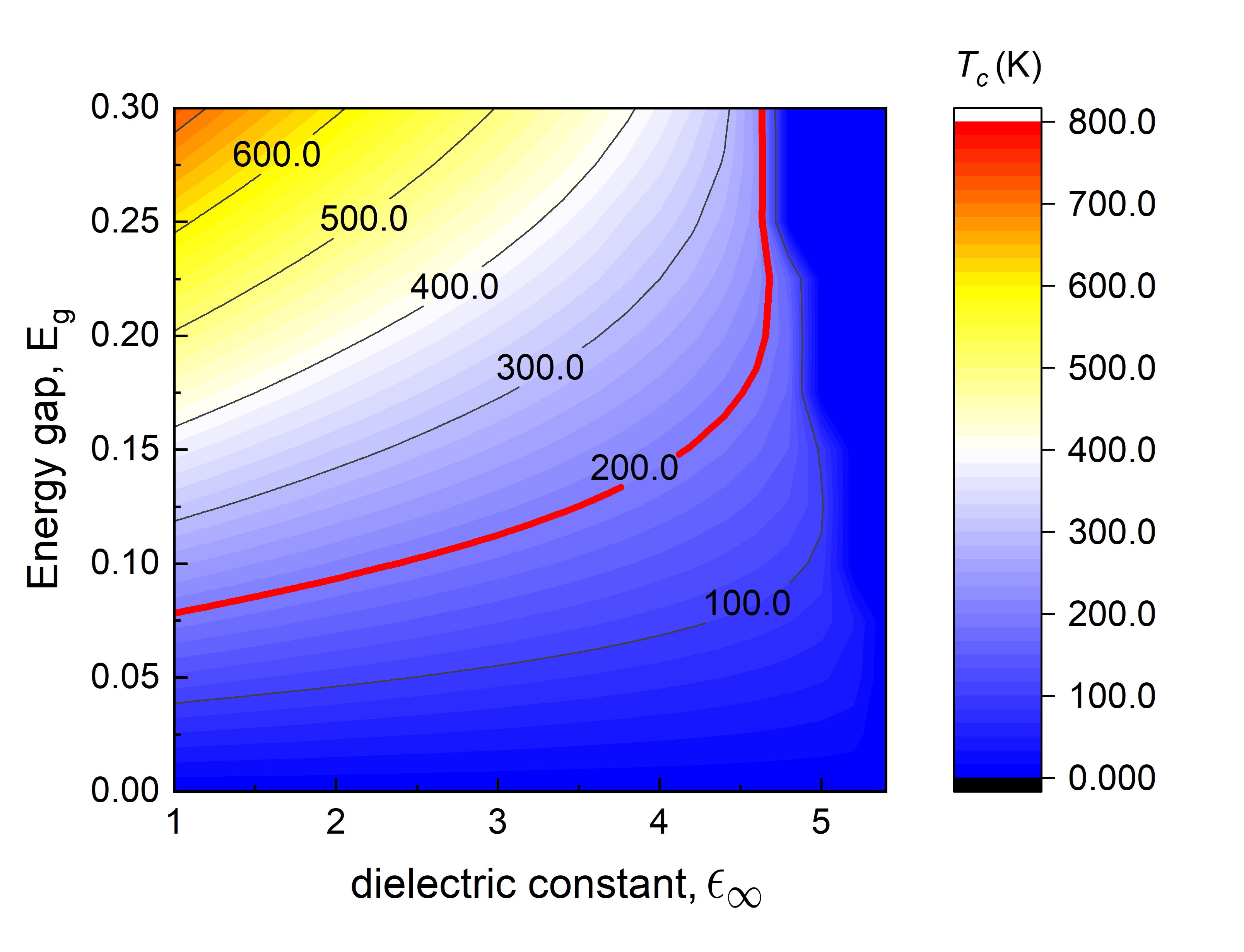}
\caption{Contour plot of the exciton critical temperature for TiSe$_2$ in the plane of the background dielectric constant $\epsilon_\infty$ and the normal gap $E_g$. The red line represents the region corresponding to $T_c = 200$ K. }\label{S_phase}
\end{figure*}

The $E_g$ where $T_c \ra 0$ on the positive $E_g$ side is the exciton binding energy $E_X $. Fig.\ \ref{S_Tc_model} shows that the $E_X$ is reduced in the indirect gap cases as given in Eq.\ (\ref{EX}) in the main text. Consider the gap equation of Eq.\ (\ref{eq:linear}). Take the $T \ra 0$ limit and determine the positive $E_g$ that satisfies the equation.
From Eq.\ (\ref{eq:baredispersion})
 \ba
\xi_e -\xi_h&=& E_g + \left( \frac{\hbar^2}{2m_e} +\frac{\hbar^2}{2m_h} \right) k^2
 +\frac{\hbar^2}{2m_e} \left( \overline{\Gamma M}^2 -2k \overline{\Gamma M} \right) \nonumber \\
&=& E_g + \frac{\hbar^2 \overline{\Gamma M}^2}{2M}
 +\frac{\hbar^2}{2m_r} \left( k- \frac{m_r}{m_e} {\overline{\Gamma M}} \right)^2,
 \ea
where $M $ and $m_r$ are the total and reduced masses of an electron and a hole of an exciton,
 \ba
 M=m_e +m_h , \nonumber \\
 \frac{1}{m_r} = \frac{1}{m_e} +\frac{1}{m_h}.
 \ea
For a rough estimate of $E_X$, we neglect the $k$ dependence of the order parameter for simplicity and extend the range of integration of Eq.\ (\ref{eq:linear}) to the entire wave-vector space to obtain
 \ba
1 = \int d^2k \frac{e^2}{2\pi \epsilon_\infty} \frac{1}{k} \frac{1}{E_g + \frac{\hbar^2 \overline{\Gamma M}^2}{2M} +\frac{\hbar^2}{2m_r} k^2} \nonumber \\
= \frac{\alpha}{\epsilon_\infty} \frac{2 m_r c\pi} {\hbar} \frac{1}{\sqrt{\frac{2m_r}{\hbar^2} \left( E_g + \frac{\hbar^2 \overline{\Gamma M}^2}{2M}\right) }}
 \ea
Then, the exciton binding energy is given by
 \ba \label{S:EX}
E_X \approx 2 m_r c^2 \left( \frac{\alpha}{\epsilon_\infty} \right)^2 -\frac{\hbar^2 \overline{\Gamma M}^2}{2 M},
 \ea
as given in Eq.\ (\ref{EX}) in the main text. The first term in the right hand side is the well known binding energy of a direct exciton in 2D.
For $T_c$ of layered structures, $E_X$ is modified to
 \ba \label{S:EX2}
E_X \approx 2 m_r c^2 \frac{\alpha^2}{\epsilon_\infty^\parallel \epsilon_\infty^\perp} -\frac{\hbar^2 \overline{\Gamma L}^2}{2 M},
 \ea
$E_X$ is reduced by the indirectness of the gap. A slight increase of $\epsilon_\infty$ or a decrease of $M$ such that the first term becomes smaller than the second one in the righthand side of Eq.\ (\ref{S:EX}) then completely suppresses an EI phase in candidate materials. Fig.\ \ref{S_phase} shows a contour plot of excitonic $T_c$ of TiSe$_2$ in the plane of $\epsilon_\infty - E_g$ plane. For example, for $\epsilon_\infty \approx 4.0$, $E_X \approx 0.6$ eV, and for $\epsilon_\infty \gtrsim 4.5$ the EI phase is completely suppressed because $E_g \gtrsim  E_X$ as can be seen from Fig.\ \ref{S_phase}.

\begin{figure*}
\includegraphics[width=0.6\textwidth]{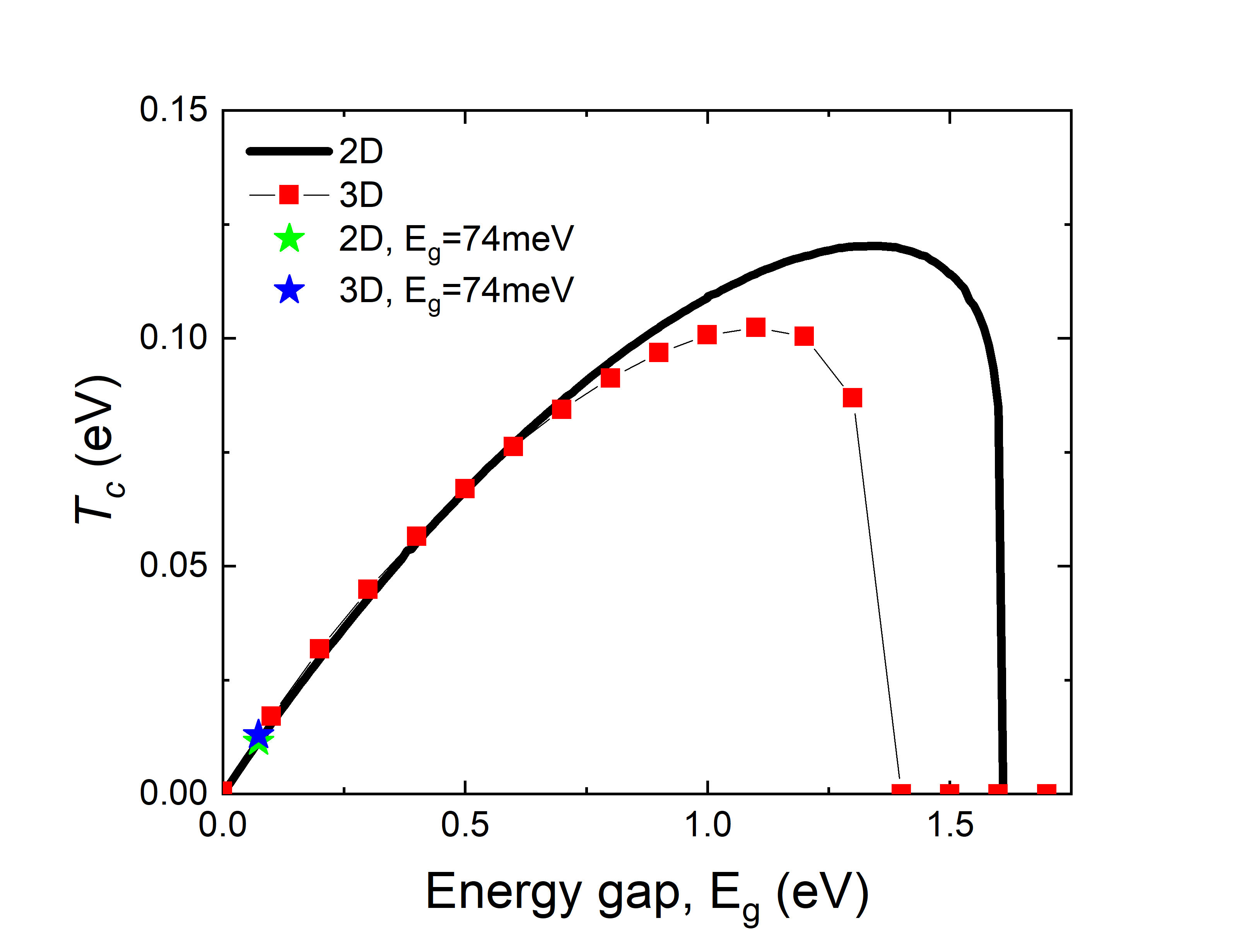}
\caption{The exciton critical temperature for TiSe$_2$ as a function of the normal gap $E_g$ from 3D gap equation. }\label{fig:Tc_3D}
\end{figure*}

Figure \ref{fig:Tc_3D} shows the calculated critical temperature $T_c$ as a function of $E_g$ from the 3D gap equation Eq.\ (\ref{eq:linear3D}) with the screened Coulomb interaction of layered structures Eq.\ (\ref{eq:Coulomb3D}) in comparison with the 2D results from Fig.\ \ref{phasediag}{(a)}. The $E_g$ refers to the minimum gap between the electron and hole bands and corresponds to the gap between $\Gamma$ and $L$ in 3D ($M$ in 2D). The $k_z$ dispersion, therefore, widens the gap between the electron and hole bands on average which around 74 meV negates weakening of screened Coulomb interaction and leads to an increases of $T_c$.
We adapted the parameterization for the $k_z$ dispersion as given by Monney $et~al$.\cite{Monney2009} adding
 \ba 
t_e & & \cos \left( \pi k_z / \overline{\Gamma A} \right), \nonumber \\
t_h & & \cos \left( \pi k_z / \overline{\Gamma A} \right) 
 \ea
to the electron and hole dispersions, respectively, with $t_e = 30 \pm 2.5$ meV and $t_h = 60 \pm 5$ meV.
For $\epsilon_\infty^\perp = 3.25$ as determined from the {\it ab initio} calculation as explained in Appendix \ref{sec:epsilon} and $\epsilon_\infty^\parallel = 3.0$ the same as the 2D calculations, $T_c$ increased to 150 K around $E_g \approx $ 74 meV as shown in Fig.\ \ref{fig:Tc_3D}. Another change is that the exciton binding energy $E_X$ decreased to around 1.3 eV in 3D (1.6 in 2D). This is as anticipated because the first term of (\ref{S:EX2}) in comparison with (\ref{S:EX}) decreases and the second term increases in 3D.

\begin{figure*}
 \includegraphics[width=0.7\linewidth]{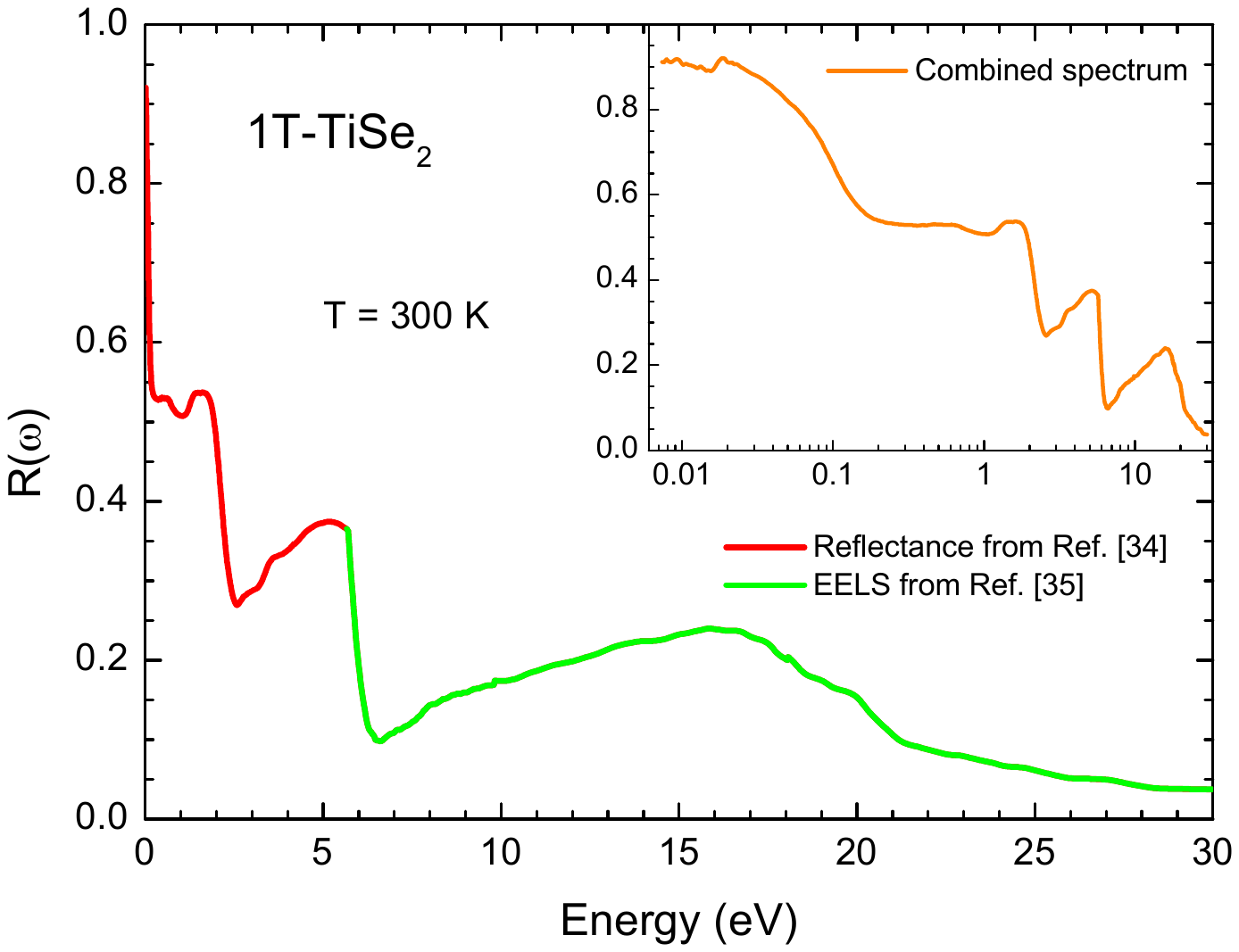}
\caption{Separate reflectance spectra of 1$T$-TiSe$_{2}$ at 300 K obtained from the optical spectroscopy\cite{Li2007} and the EELS study.\cite{Shu2019} In the inset we show the combined reflectance.}
    \label{spectrum}
\end{figure*}

\begin{figure*}
 \includegraphics[width=0.7\linewidth]{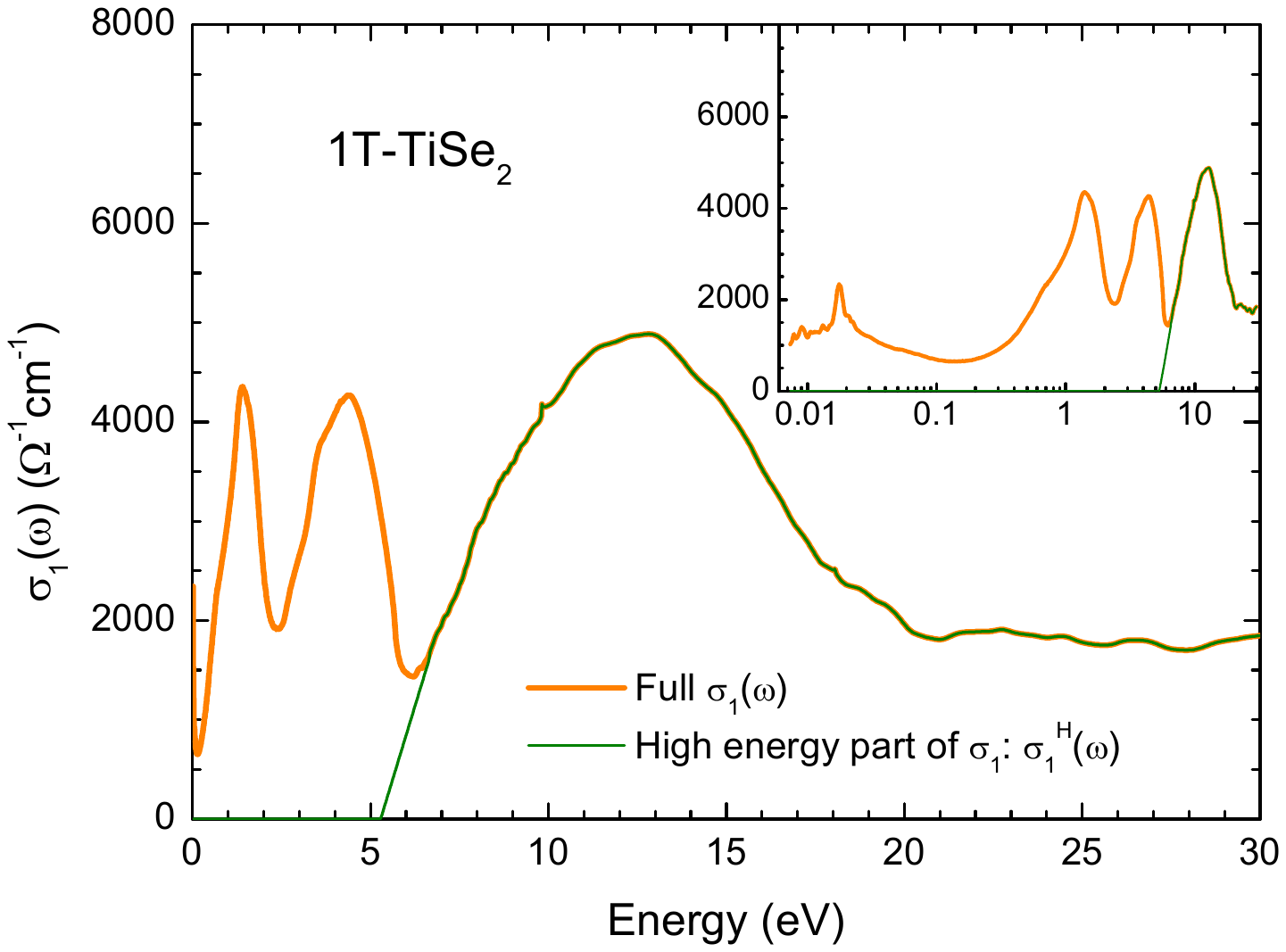}
\caption{The optical conductivity: the full conductivity in the orange line and a high energy part above 6.5 eV in the olive line.}
    \label{optical}
\end{figure*}

\section{Determination of ${\boldsymbol{\varepsilon }}_{\boldsymbol{\infty }}$ of 1$T$-TiSe$_2$
  \label{sec:epsilon} }

We determined the background dielectric constant ${\varepsilon }_{\infty }$ by combining the measured reflectance spectrum below 6.2 eV,\cite{Li2007} and the measured electron energy loss spectroscopy spectrum up to 30 eV.\cite{Shu2019} We also used the {\it ab initio} calculated real and imaginary parts of the dielectric functions in a published paper\cite{Leventi1995} to check reliability of the determined value from the experiments. They are in good agreement with each other as we discussed below.

The reflectance spectrum $R(\omega)$ for the normal incidence is given in terms of the complex dielectric constant $\widetilde{\varepsilon }\left(\omega \right)\equiv {\varepsilon }_1(\omega )+{i\varepsilon }_2(\omega )$ by
 \ba
R\left(\omega \right)= \left| \frac{1-\sqrt{\widetilde{\varepsilon }\left(\omega \right)} } {1+\sqrt{\widetilde{\varepsilon }\left(\omega \right)} }\right|^2 .
  \ea
The separate reflectance spectra at 300 K are shown in Fig.\ \ref{spectrum}, and the combined reflectance spectrum is shown in the inset. We performed the Kramers-Krong (KK) analysis \cite{WootenBOOK} to get the optical conductivity from the combined reflectance spectrum. The optical conductivity is shown in Fig.\ \ref{optical}.

\begin{figure*}
 \includegraphics[width=0.7\linewidth]{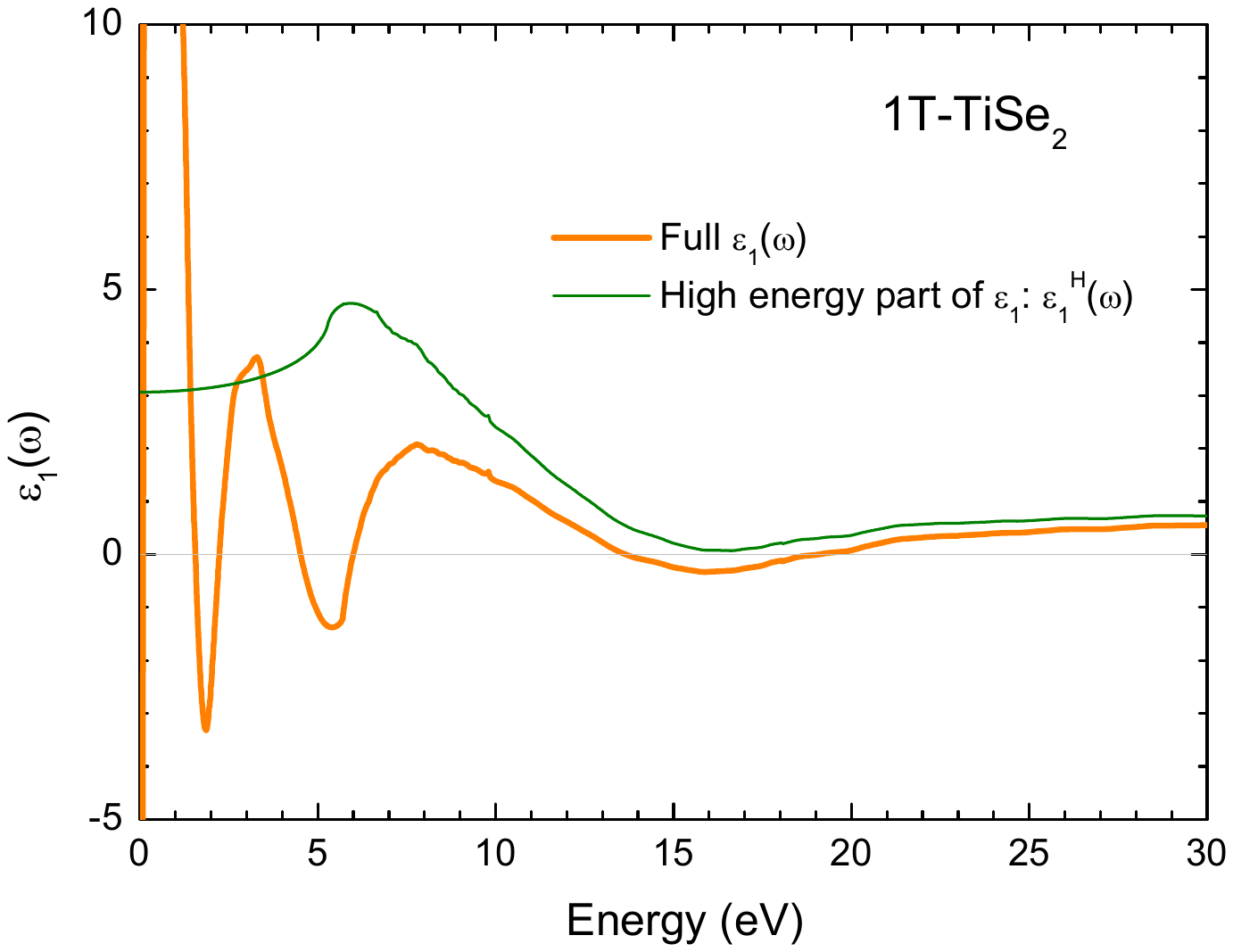}
\caption{ The dielectric function $\epsilon_1 (\omega )$ of 1$T$-TiSe$_{2}$ obtained from the KK analysis and the high energy dielectric function, $\epsilon^{H}_1 (\omega )$.}
 \label{dielectric}
\end{figure*}

\begin{figure*}
\includegraphics[width=0.6\linewidth]{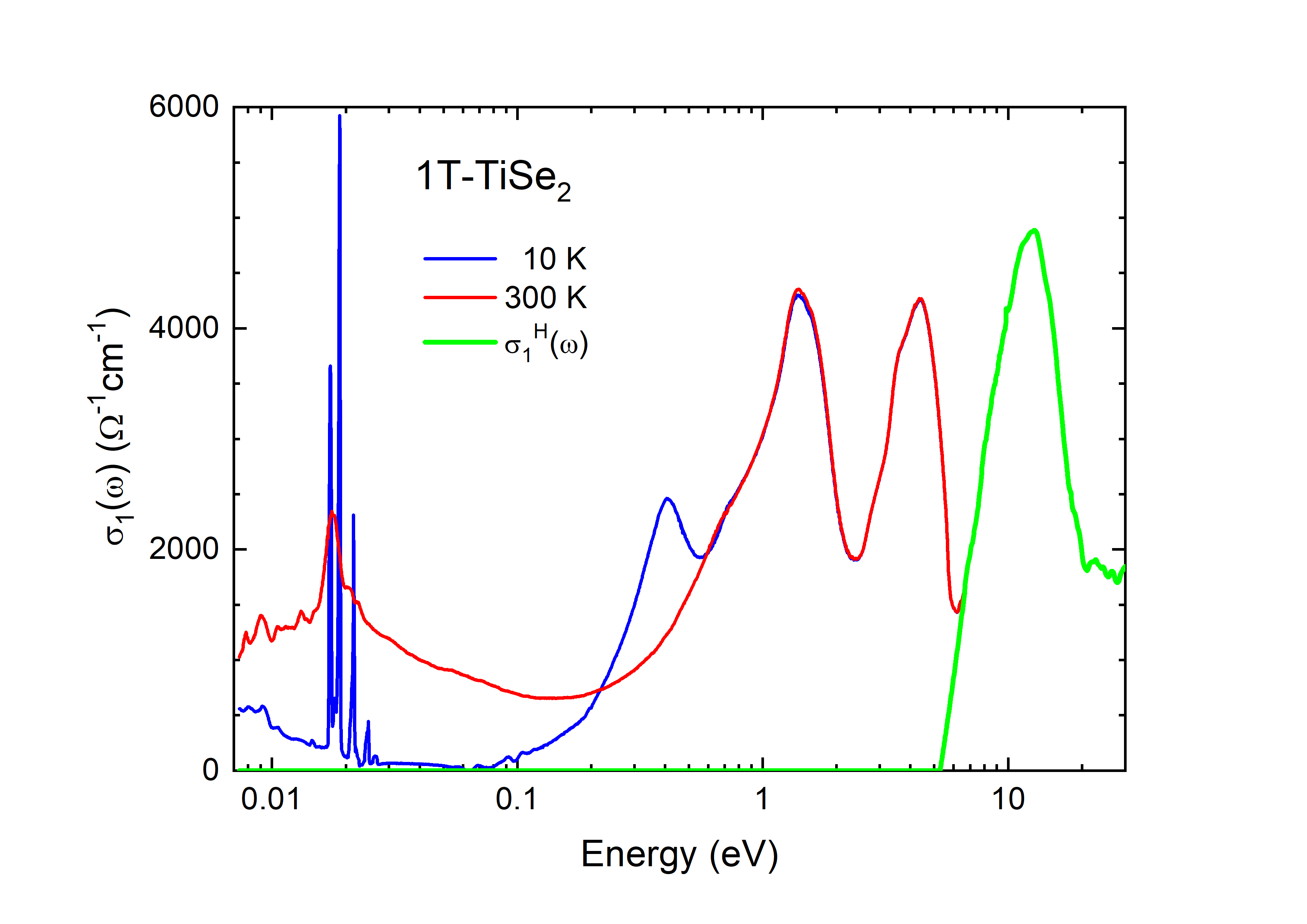}
\caption{The conductivity $\sigma_1(\omega)$ and the high energy part conductivity $\sigma_1^H (\omega)$ for the temperature 10 K and 300 K. The $\sigma_1^H (\omega)$ for the two temperatures overlap exactly and, consequently, the background dielectric constant $\epsilon_\infty$ is temperature independent.}
 \label{dielectric_lowT}
\end{figure*}

\begin{figure*}
\includegraphics[width=0.6\linewidth]{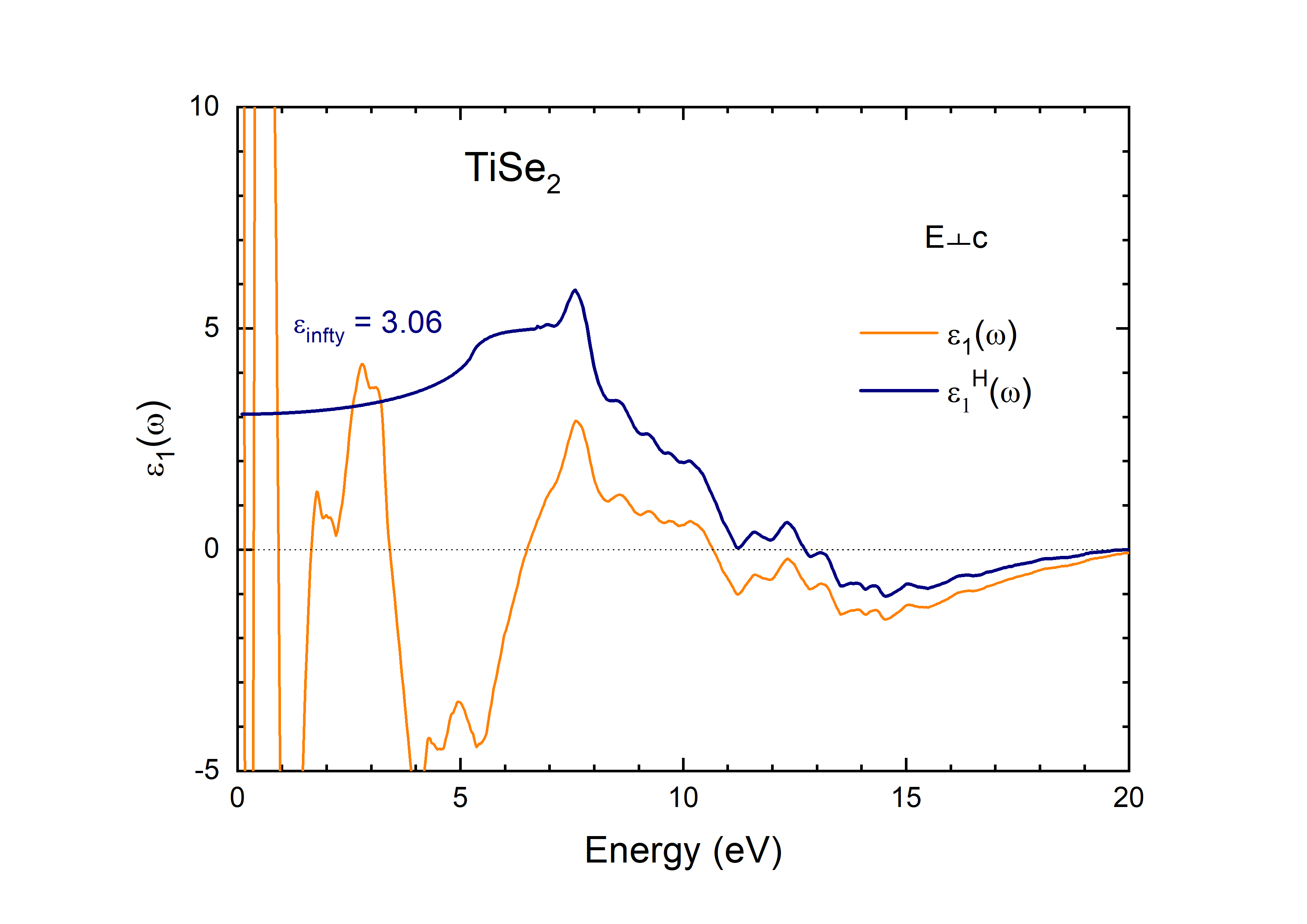}
\caption{The dielectric function $\epsilon_1 (\omega)$ and the high energy part dielectric function $\epsilon_1^H (\omega)$ for the electric field parallel to the layers. }
 \label{dielectric_parallel}
\end{figure*}

\begin{figure*}
\includegraphics[width=0.6\linewidth]{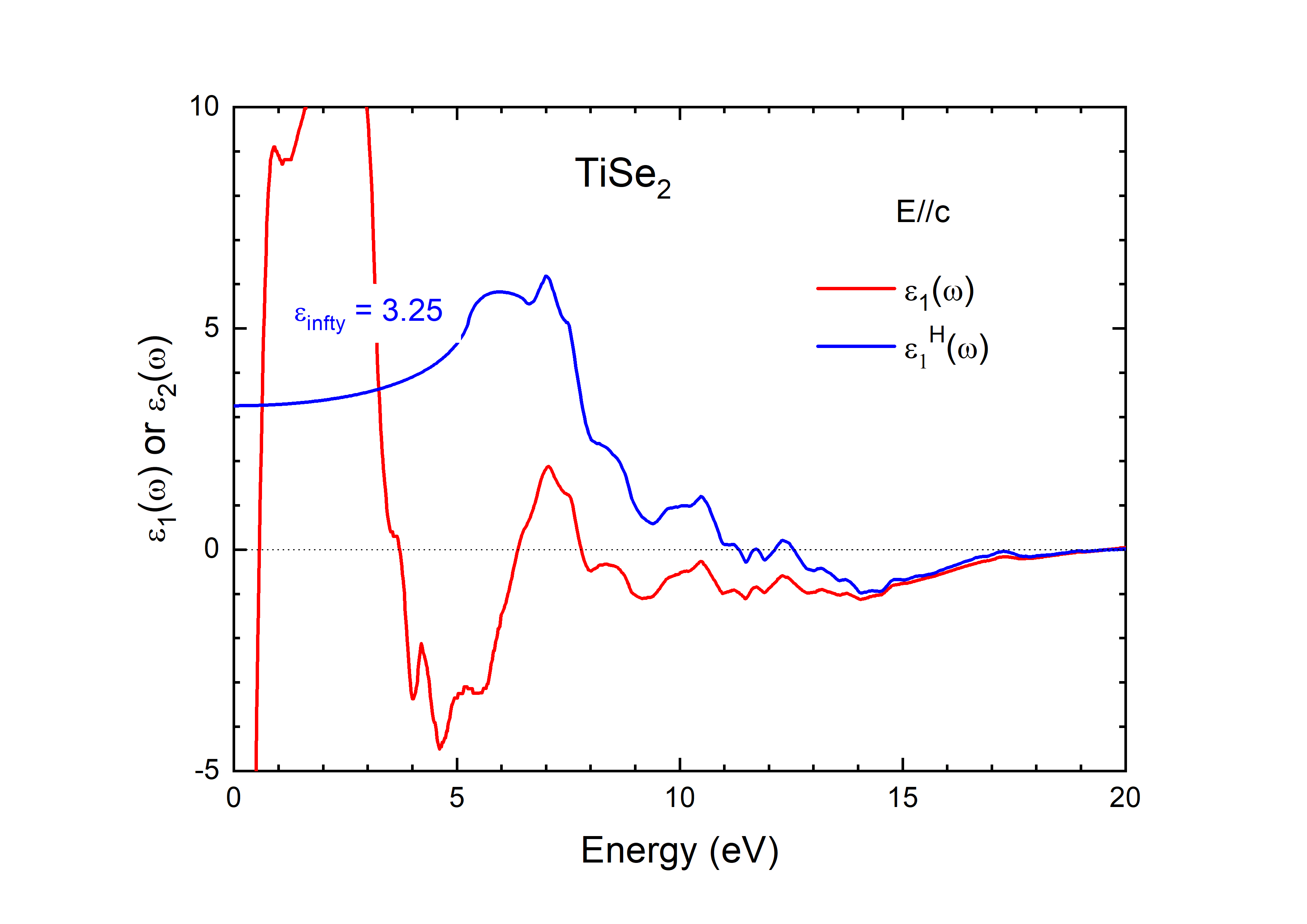}
\caption{The full dielectric function and the high energy part dielectric function for the electric field perpendicular to the layers. }
 \label{dielectric_perp}
\end{figure*}

The largest interband optical transition between the electron and hole bands of Eqs.\ (\ref{eq:baredispersion}) considered in the present calculations is about 6.5 eV within the 1st BZ. The excitations above 6.5 eV then contribute to the background dielectric constant ${\varepsilon }_{\infty }$.
We used the method introduced in \cite{Hwang2007} to estimate ${\varepsilon }_{\infty }$ from the optical conductivity above 6.5 eV. We obtain the imaginary part of the optical conductivity above 6.5 eV using the Kramers-Kronig relation \cite{WootenBOOK} between the real and imaginary parts of the optical conductivity. Also using the relation between the imaginary part of the optical conductivity and the real part of the dielectric function, we obtain the background dielectric constant as
 \ba
{\varepsilon }_{\infty }\equiv {\mathop{\mathrm{lim}}_{\omega \to 0} {\varepsilon }^H_1\left(\omega \right)\ } , \nonumber \\
{\varepsilon }^H_1\left(\omega \right)=1-\ \frac{4\pi }{\omega }{\sigma }^H_2\left(\omega \right).
 \ea
${\varepsilon }_{\infty } = 3.0 $ as can be seen in Fig.\ \ref{dielectric}.

The ${\varepsilon }_{\infty }$, having been determined by the contributions from high-energy excitations above the cutoff of 6.5 eV, should be temperature independent. To demonstrate this explicitly, we used the 10 K experimental data to determine ${\varepsilon }_{\infty }$ at $T=10$ K. As can be seen from the Fig.\ \ref{dielectric_lowT} the conductivity at 10 K (blue curve) and 300 K (red) merge above around 1 eV. The ${\varepsilon }_{\infty }$ is determined from the high energy part (green curve) where the two different $T$ data overlap exactly. This explicitly shows that the background dielectric constant ${\varepsilon }_{\infty }$ is temperature independent as expected.

Since the exciton instability is rather sensitive on the $\epsilon_\infty$ value as can be seen, for example, from Eq.\ (\ref{EX}), we checked the reliability of the determined value of ${\varepsilon }_{\infty } = 3.0 $. We proceeded exactly the same now with the {\it ab initio} evaluated real and imaginary parts of the dielectric functions $\epsilon_1 (\omega)$ and $\epsilon_2 (\omega)$ for the inplane electric field (${\bf E} \perp {\hat z} $) in a published paper.\cite{Leventi1995}
We obtained ${\varepsilon }_{\infty }^\parallel = 3.06 $ as can be seen in Fig.\ \ref{dielectric_parallel} in good agreement with the ${\varepsilon }_{\infty }^\parallel = 3.0 $ from the combined data of reflectance and EELS.

We also determined ${\varepsilon }_{\infty }^\perp $ from the {\it ab initio} calculations for the out-of-plane electric field.\cite{Leventi1995} We obtained ${\varepsilon }_{\infty }^\perp = 3.25 $ as can be seen in Fig\ \ref{dielectric_perp}. Then, the 3D effective background dielectric constant is $\sqrt{ {\varepsilon }_{\infty }^\parallel {\varepsilon }_{\infty }^\perp} = 3.15 $.

\section{Comparison with STS/STM experiments  \label{sec:dos}}

\begin{figure*}
\includegraphics[width=0.5\linewidth]{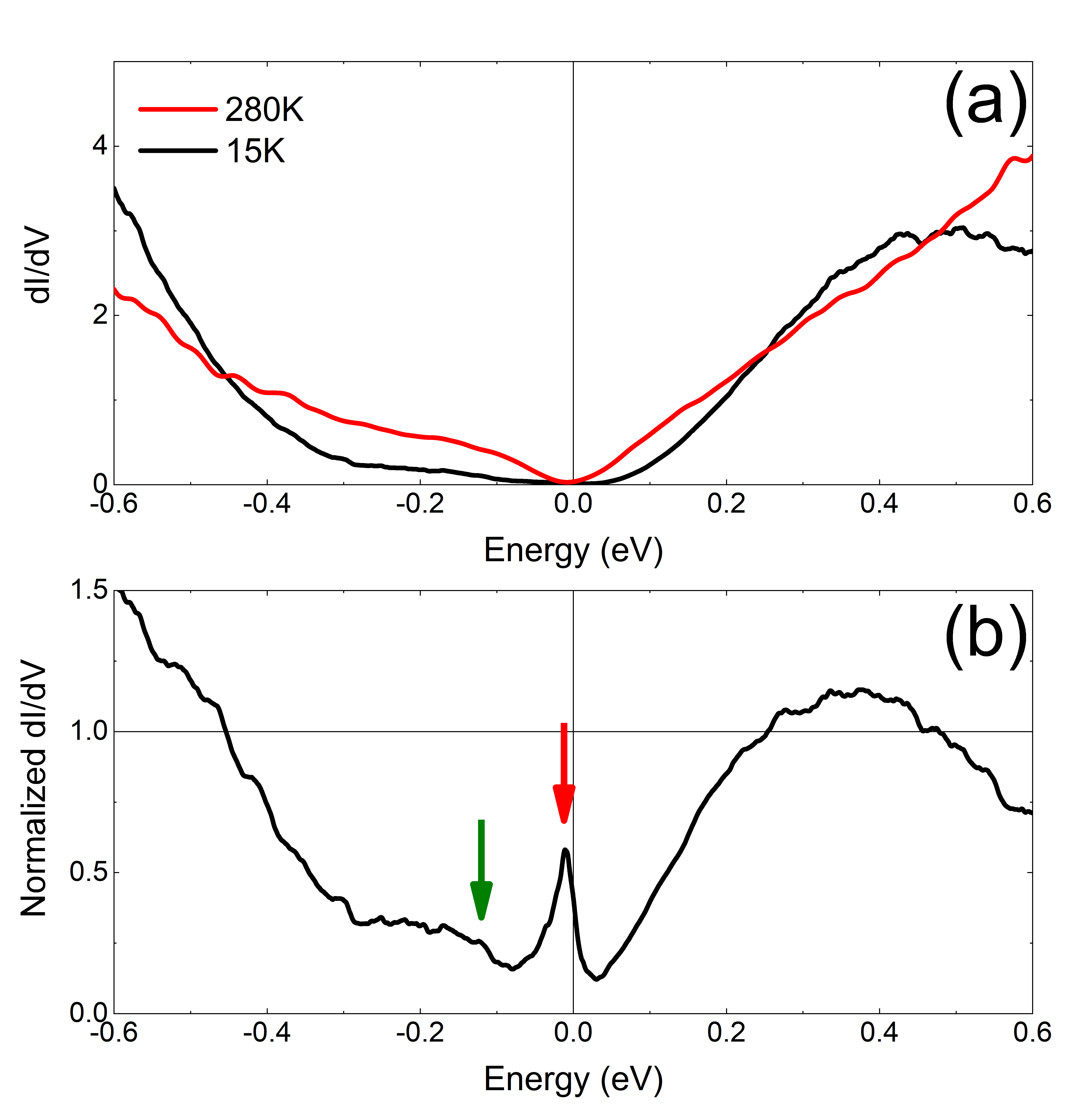}
\caption{(a) The dI/dV of TiSe$_{2}$ 280K (red) and 15K (black) from Kolekar $\textit{et al}$  \cite{Kolekar2018}. (b) Normalized dI/dV. }
\label{sdos}
\end{figure*}

In the low temperature limit, we need to solve the full gap equation of Eq.\ (\ref{eq:gap}) in the main text. The ${\bf k}'$ summation was performed using the fast Fourier transform (FFT) between the momentum and real space for accuracy and efficiency using the convolution relation
 \ba
\sum_{\bf k} e^{i {\bf k}\cdot{\bf r}} \sum_{{\bf k}'} F({\bf k}'-{\bf k}) G({\bf k}') =
F({\bf r}) G({\bf r}).
 \ea
Then, Eq.\ (\ref{eq:gap}) may be transformed to the real space as
 \ba
 \label{eq:gap_realspace}
\Delta_{i}(\mathbf{r}) = -\int_{-\infty}^{\infty}d\omega A_{\Delta_{i}}(\mathbf{r},\omega)f(\omega)V_{s}(\mathbf{r}).
 \ea
The screened Coulomb interaction and the spectral function in the real space $V_s (\mathbf{r})$ and $A(\mathbf{r},\omega)$, respectively, were also obtained by performing FFT.
The $k$-points in the 2D hexagonal lattice (gray area in Fig.\ \ref{k_points}) are not suitable for performing FFT, so, a 2D rhombus lattice (dashed in the figure) of the same size was considered.
The $k$-points were selected evenly spaced within the 2D rhombus lattice (red and black dots in the figure).
The selected $k$-points naturally include the periodicity of the 2D hexagonal lattice.
At this time, in order to fully consider the information of the 2D hexagonal lattice, the number of $k$-points on one side was set to an integer multiple of 6. Here we choose $48 \times 48$ $k$-points as shown in Fig.\ \ref{k_points}.
Since the rhombus is an inclined rectangle, the selected $k$-points can be applied to the conventional 2D FFT algorithm.

The gap equation of (\ref{eq:gap_realspace}) was solved self-consistently via numerical iterations. The $\Delta_i ({\bf k})$ $(i=1,2,3)$ are all coupled in the full gap equation. They were obtained simultaneously from iterations without making symmetry operations among them. The obtained $\Delta_i ({\bf k})$ were shown in Fig.\ \ref{Delta}.
With thus determined $\Delta_i ({\bf k})$ the spectoscopic features were calculated as explained in main text.
The semiconducting DOS of our model TiSe$_2$ (Fig.\ \ref{arpes}(c) red solid line in the main text) gets enlarged in EI phase (Fig.\ \ref{arpes}(c) black solid line in the main text). Some detailed structures, split weak coherence peaks and step-like structures are as we may expect from the ARPES intensity.

These features may be revealed more clearly in the normalized DOS as shown in Fig.\ \ref{arpes}(d) in the main text. A remarkable structure in the normalized DOS from calculations is the zero bias peak,
which is caused by semiconducting normal state.
The corresponding DOS plots from STS experiments by Kolekar \textit{et al} \cite{Kolekar2018} are shown for both normal (Fig.\ \ref{sdos}(a) red solid line) and CDW phase (Fig.\ \ref{sdos}(a) black solid line), and the normalized DOS in Fig.\ \ref{sdos}(b).
It is not straightforward to determine the gap size from experiments because there is no strong coherence peak.
But this weak coherence peak is consistent with calculations.
A step-like structure around $-0.1 \sim -0.2$ eV can be compared with calculations, although the energy scale is not clear.
Interestingly, the zero bias peak also appears in normalized DOS from experiments.
For more systematic comparison, one may need a systematic measurement and analysis of STS/STM for above and below $T_c$.


\vspace{1cm}

\bibliographystyle{unsrt}

\bibliography{references_v2}

\vspace{1cm}

{\bf Acknowledgements}

We thank Sadhu Kolekar and Matthias Batzill for providing us with the experimental STS data, and Matthew Watson, Claude Monney, Chandra Varma, and Yunkyu Bang for discussion and comments on the manuscript.
The work was supported by the Samsung Science and Technology Foundation (SSTF) through Grant No.\ SSTF-BA1502-06 and by Nation Research Foundation (NRF) of Korea through Grant No.\ NRF-2018R1D1A1B07043997 (HYC), NRF-2019R1I1A1A01057393 (JMB), and NRF-2019R1A6A1007307912 (JH). Correspondence and requests for materials should be addressed to HYC (e-mail: hychoi@skku.edu).

\end{document}